\documentclass[%
 reprint,
superscriptaddress,
 amsmath,amssymb,
 aps,
]{revtex4-2}

\usepackage{graphicx}
\usepackage{dcolumn}
\usepackage{bm}
\usepackage{mathtools}
\usepackage{braket}
\usepackage{physics}
\usepackage[utf8]{inputenc}
\usepackage{tikz}\usetikzlibrary{quantikz}
\usepackage{bbm}
\usepackage{xcolor}

\DeclareMathOperator*{\average}{average} 
\newcommand{\ketbraind}[3]{\ket{#1}_#2\! \bra{#3}}
\usepackage{subfigure}

\usepackage{layout}

\begin{document}
\preprint{APS/123-QED}

\title{Nonunitary Gate Operations by Dissipation Engineering}

\author{Elias Zapusek}
\email{zapuseke@ethz.ch}
\affiliation{
Institute for Quantum Electronics, ETH Z\"urich, 8093 Z\"urich, Switzerland
}

\author{Alisa Javadi}
\affiliation{
Department of Physics, University of Basel, Klingelbergstrasse 82, CH-4056 Basel, Switzerland
}

\author{Florentin Reiter}%
\email{freiter@phys.ethz.ch}
\affiliation{
Institute for Quantum Electronics, ETH Z\"urich, 8093 Z\"urich, Switzerland
}

\date{\today}

\begin{abstract}

Irreversible logic is at odds with unitary quantum evolution. Emulating such operations by classical measurements can result in disturbances and high resource demands. To overcome these limitations, we propose protocols that harness dissipation to realize the nonunitary evolution required for irreversible gate operations. Using additional excited states subject to decay, we engineer effective decay processes that perform the desired gate operations on the smallest stable Hilbert space. These operate deterministically and in an autonomous fashion, without the need for measurements. We exemplify our approach considering several classical logic operations, such as the OR, NOR, and XOR gates. Towards experimental realization, we discuss a possible implementation in quantum dots. Our study shows that irreversible logic operations can be efficiently performed on realistic quantum systems and that dissipation engineering is an essential tool for obtaining nonunitary evolutions. The proposed operations expand the quantum engineers' toolbox and have promising applications in NISQ algorithms and quantum machine learning.

\end{abstract}

\maketitle


\section{\label{sec:intro}Introduction}

The rapid miniaturization of the elementary device of classical computation, the transistor, has enabled the information age. These days, however, silicon transistors are quickly approaching their fundamental scaling limit \cite{noauthor_international_2015}. This has prompted the exploration of a variety of different technologies to continue the increase in computational power \cite{shulaker_carbon_2013, ma_small-diameter_2003, han_energy_2007, datta_electronic_1990, sugahara_spin_2004, salahuddin_use_2008, chen_integrated_2008, zhou_field_1997, wray_topological_2012, lee_conceptual_2008, andreakou_optically_2014, feng_room_2005, sharma_compact_2018, currivan-incorvia_logic_2016, amaru_majority_2018}. 
In parallel, the field of quantum information processing has developed \cite{nielsen_quantum_2010}. Here, to achieve the desired quantum effects, systems are often inherently miniaturized, with individual atoms \cite{monroe_demonstration_1995, benhelm_towards_2008, harty_high-fidelity_2014, mehta_integrated_2020, jaksch_fast_2000, levine_high-fidelity_2018} or solid-state structures \cite{chow_universal_2012, krinner_realizing_2021, loss_quantum_1998,imamoglu_quantum_1999, nichol_high-fidelity_2017} carrying the information.

\begin{figure}[t!]
    \centering
    \includegraphics[scale=0.5]{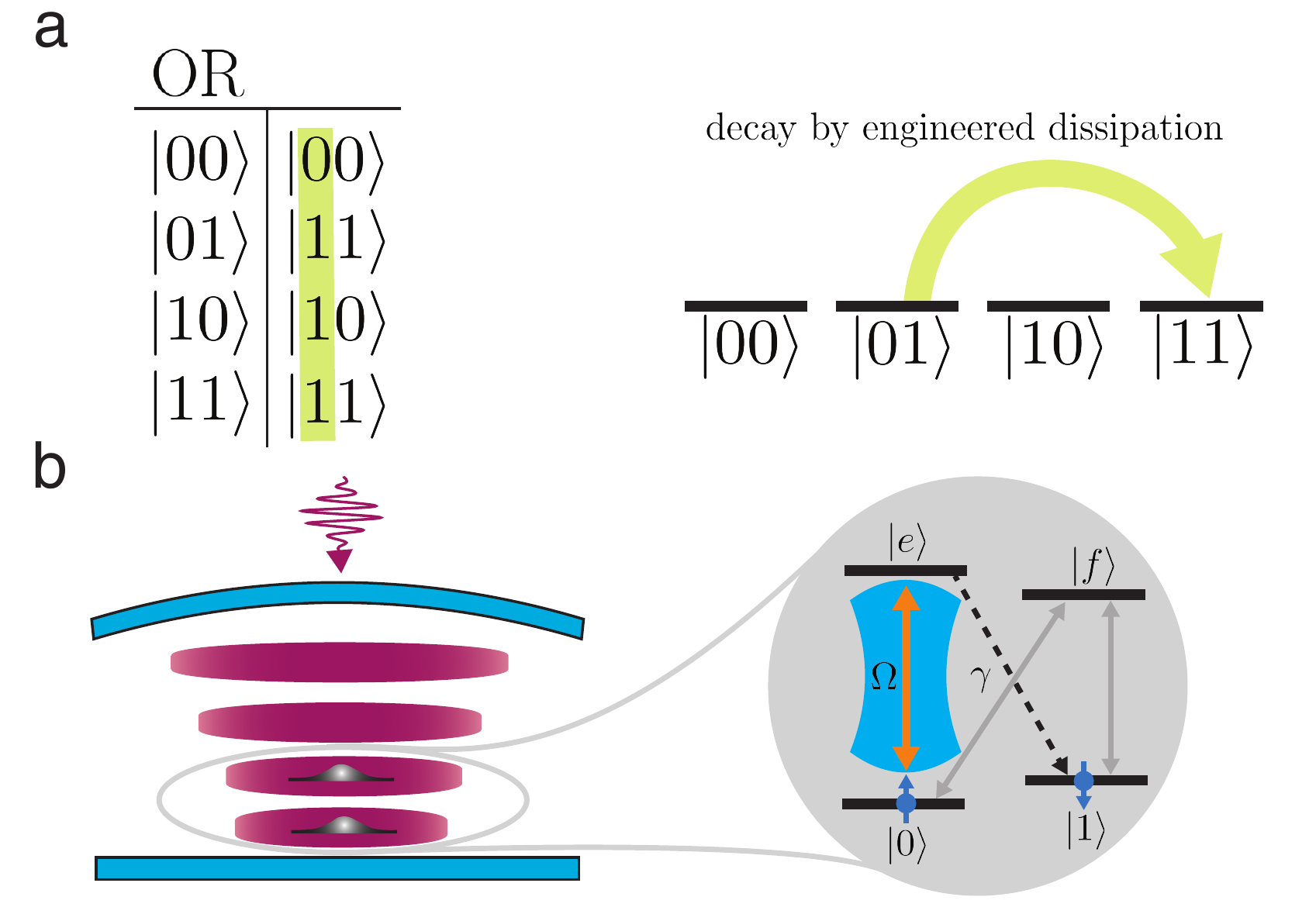}
    \caption{\textbf{Nonunitary Gate Operation} (a) Truth table of the OR gate. The logical output is highlighted in green. We choose to display the result on the first qubit, therefore we need to engineer the mapping $\ket{01}\rightarrow\ket{11}$. (b) Two quantum dots in the strong coupling regime in a microcavity. Their level scheme is shown on the right. The ground states are used for the logical states while the excited state $\ket{e}$ is used during gate operation. The quantum dots are subject to coherent couplings (solid lines) and dissipation (dotted lines).}
    \label{fig:idea}
\end{figure}

From the combination of the paradigms of classical and quantum information processing stems the idea to perform classical logic operations in quantum systems \cite{terashima_nonunitary_2005, wang_autonomous_2019, chu_scalable_2021}. 
The term \textit{logic gate} commonly refers to an electronic circuit that implements a Boolean function. This Boolean function satisfies a truth table (Fig.\,\ref{fig:idea}(a)) containing all input-output pairs \cite{holdsworth_digital_2002}. Any quantum evolution that realizes a type of gate thus needs to reproduce the truth table.
This is challenging as, e.g. in classical gates, the functions are irreversible, and consequently also nonunitary. Since quantum computation is largely realized by unitary operations \cite{cirac_quantum_1995,molmer_multiparticle_1999,paraoanu_microwave-induced_2006}, irreversible dynamics, such as decoherence, are strictly avoided \cite{divincenzo_physical_2000,ladd_quantum_2010}.

Nonunitary operations have recently gained attention in the light of near-term quantum algorithms performed on Noisy Intermediate-Scale Quantum (NISQ) devices \cite{preskill_quantum_2018, bharti_noisy_2021,mazzola_nonunitary_2019,del_re_driven-dissipative_2020, hu_quantum_2020, hu_general_2021,ramusat_quantum_2021, wang_quantum_2011, verdon_quantum_2019,zhu_generation_2020,wu_variational_2019, wang_variational_2021,foldager_noise-assisted_2021,zoufal_variational_2021,     verdon_quantum_2019-1,   lee_neural-network_2021, cong_quantum_2019,herrmann_realizing_2021}:

To realize them, various approaches have been suggested for tasks such as quantum optimization \cite{mazzola_nonunitary_2019,wang_variational_2021}, quantum simulation \cite{del_re_driven-dissipative_2020, hu_quantum_2020, hu_general_2021,ramusat_quantum_2021, wang_quantum_2011, verdon_quantum_2019,zhu_generation_2020,wu_variational_2019, wang_variational_2021,foldager_noise-assisted_2021}, and quantum machine learning \cite{wang_variational_2021,foldager_noise-assisted_2021,zoufal_variational_2021,     verdon_quantum_2019-1,   lee_neural-network_2021, cong_quantum_2019,herrmann_realizing_2021}. Nonunitary operations are used for the generation of low temperature thermal states \cite{foldager_noise-assisted_2021, wu_variational_2019,zhu_generation_2020,wang_variational_2021}, as a projective filter \cite{mazzola_nonunitary_2019}, to simulate open quantum systems \cite{del_re_driven-dissipative_2020, hu_quantum_2020, hu_general_2021,ramusat_quantum_2021,lee_neural-network_2021}, or in pooling layers of quantum neural networks \cite{cong_quantum_2019, herrmann_realizing_2021}. To a large extent, the available realizations rely on measurements of (ancilla) qubits and conditional operations. The requirement for macroscopic measurement devices and feedforward can lead, however, to high resource overheads, as well as imperfections that are challenging with current devices \cite{chu_scalable_2021, elder_high-fidelity_2020}.

Specifically, the challenge of reconciling irreversible logic with quantum mechanics can be tackled in a number of different ways \cite{bennett_logical_1973, goos_reversible_1980, monz_realization_2009, terashima_nonunitary_2005, williams_probabilistic_2004, daskin_ancilla-based_2017, liu_probabilistic_2021}. First and foremost, one can construct a universal set of reversible logic gates. This, however, requires enlarging the system as for a two-qubit gate, a third qubit to store the result is needed \cite{bennett_logical_1973,goos_reversible_1980,monz_realization_2009}.
Alternatively, one can use projective measurements to construct physically irreversible gates. These gates do not require a larger system, but due to their reliance on measurements, they are probabilistic \cite{terashima_nonunitary_2005,williams_probabilistic_2004,  daskin_ancilla-based_2017, liu_probabilistic_2021}.

In contrast to these works, we use dissipation as a resource to facilitate nonunitary operations. Employing dissipation, a range of quantum information tasks can be performed, such as state preparation by optical pumping \cite{kastler_quelques_1950,wineland_optical_1985,bruzewicz_trapped-ion_2019}, squeezing \cite{kronwald_dissipative_2014,agarwal_strong_2016}, entanglement generation \cite{plenio_cavity-loss-induced_1999,vacanti_cooling_2009,kastoryano_dissipative_2011,krauter_entanglement_2011,reiter_driving_2012,lin_dissipative_2013,ticozzi_steady-state_2014,reiter_scalable_2016,lin_preparation_2016}, quantum error correction \cite{cohen_dissipation-induced_2014,reiter_dissipative_2017}, and quantum simulation \cite{diehl_quantum_2008,barreiro_open-system_2011}. This is achieved without the need for a classical measurement apparatus, and holds scaling and robustness advantages over unitary quantum information processing approaches \cite{kastoryano_dissipative_2011, morigi_dissipative_2015}.

We exemplify our approach by demonstrating the deterministic realization of classical logic gates. Specifically, we utilize the spontaneous emission of a quantum emitter to perform the action of an OR gate. Our proposed mechanism is minimal in the dimension of the required stable Hilbert space and can be implemented on different quantum emitters, ranging from atomic systems \cite{lin_dissipative_2013} to solid-state quantum emitters \cite{gonzalez-tudela_entanglement_2011}. Here, we specifically analyze the performance of our gate for quantum dots in optical microcavities which have recently gained popularity for studying cavity quantum electrodynamics \cite{laussy_strong_2008,wei_scalable_2015}.

Quantum dots have reasonably coherent spin states, while possessing coherent optical transitions \cite{bodey_optical_2019,warburton_single_2013}. Additionally, the dipole moment of the optical transitions is strong which allows fast gate operations. They also exhibit a Lambda level structure which is essential for the implementation of our gate, see Fig.\,\ref{fig:idea} (b). Finally, they have been successfully embedded in microcavities \cite{loo_optical_2012,najer_gated_2019,antoniadis_chiral_2021} and nanocavities \cite{reinhard_strongly_2012} with cooperativities exceeding $100$ which is another important factor for realization of our gate.

The paper is organized in the following way: In Sec.\,\ref{sec:Protocol} we introduce the protocol for the translation of the truth table of a gate into its dissipative implementation. We use our main example, the OR gate, to illustrate the detailed mechanisms of operations. In Sec.\,\ref{sec:Setup} the couplings necessary for the OR gate are described and physical error processes are discussed. The following Sec.\,\ref{sec:Mechanism} explains the engineered mechanisms for gate operation using analytical tools. In Sec.\,\ref{sec:Performance} the system parameters are optimized numerically in an experimentally realistic setting. Next, building from the OR gate, the realization of other gates such as AND, NOR and XOR is discussed in Sec.\,\ref{sec:FurtherGates}. Finally, in Sec.\,\ref{sec:Conclusion} we conclude our findings and discuss possible applications of our novel operations.

\section{Protocol}\label{sec:Protocol}
To realize a classical multi-qubit gate, we first decide on a qubit to display our output. Then we choose mappings for all input states, which translate them to final states with the correct value on the output qubit. Next, we engineer these mappings utilizing the effective operator formalism detailed in Ref.\,\cite{reiter_effective_2012}. After the operation is complete, we trace over the other qubits, retaining only the output, and measure the result.
We exemplify the procedure considering the classical OR gate:

The OR gate is a non-reversible two-qubit gate. It maps initial state 00 to output state 0, and all other states to 1. It is equivalent to AND under exchange of 0 and 1. The protocol from truth table to dissipative mapping is illustrated for the OR gate in Fig.\,\ref{fig:idea} (a). We choose to display the result of the operation on the first qubit. States $\ket{00}$, $\ket{10}$ and $\ket{11}$ already display the correct result on the output qubit. We thus only need to realize the mapping
\begin{align}
    \ket{01} \rightarrow \ket{11}.
    \label{eq:mapping}
\end{align}
This mapping, indicated by the arrow in Fig.\,\ref{fig:idea} (a), is realized continuously by an effective decay process.
\begin{figure}
    \centering
    \includegraphics[scale=0.5]{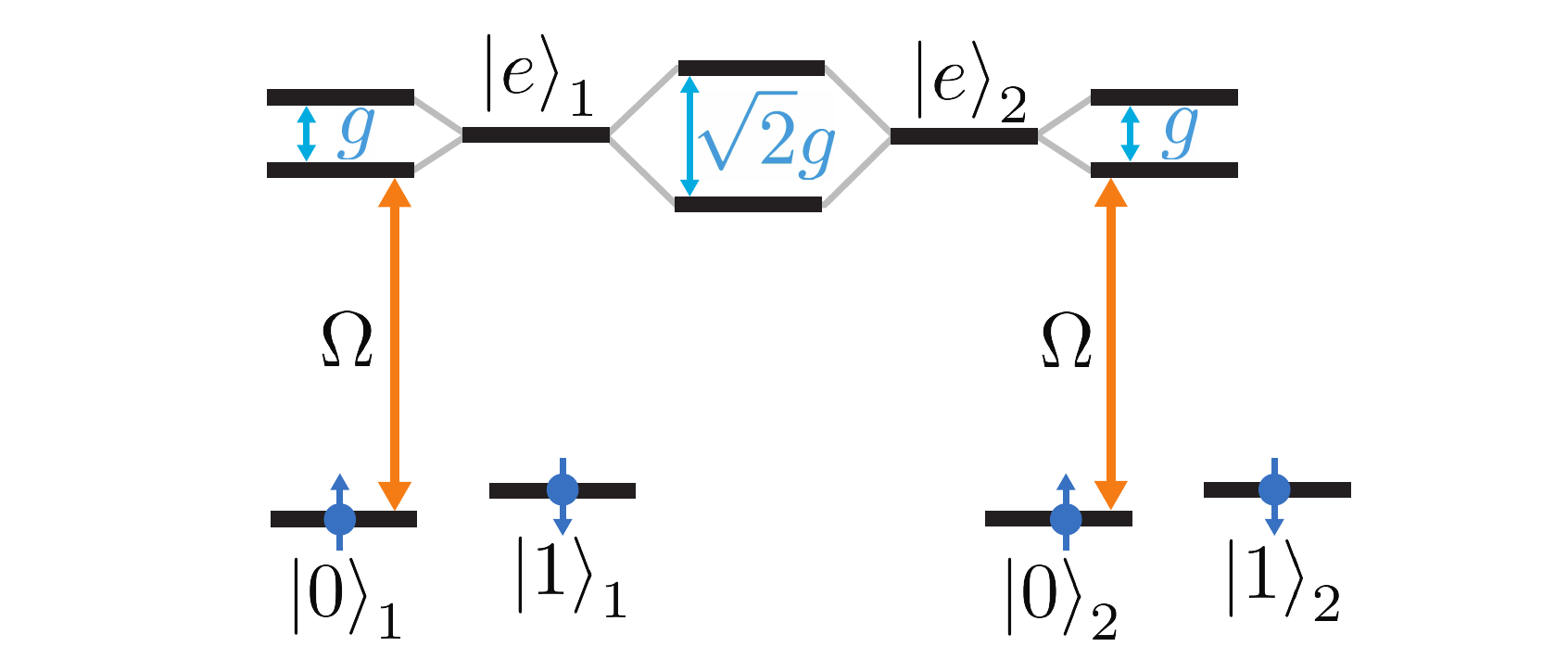}
    \caption{\textbf{Resonance Engineering.} By adjusting the detunings of the excited states only certain dressed states of the combined two-qubit system are shifted into resonance. This allows us to selectively excite states, which is the key to realizing the truth tables representing gate operations. Here, we exemplify the mechanism realizing an OR gate.}
    \label{fig:ResEngi}
\end{figure}

Fig.\,\ref{fig:ResEngi} shows the main principle used to engineer the mapping, \textit{resonance engineering}. The ground states $\ket{0}$ of both qubits are off-resonantly driven to an excited subspace. Within the excited subspace, the states interact strongly, forming dressed states. We tune the parameters such that some dressed states are shifted into resonance while others are off-resonant. The coupling of the dressed states addressed from $\ket{00}$ is enhanced to $\sqrt{2}g$. This allows us to drive only ground state $\ket{01}$ to the excited subspace. Finally, spontaneous emission completes the action of the gate.

\section{System}\label{sec:Setup}
We consider two quantum dots that strongly interact with a cavity mode. In Fig.\,\ref{fig:idea} (b) the experimental setup and level scheme are illustrated.
The system dynamics can be described by a master equation in Lindblad form
\begin{equation}\label{eq:lind}
    \Dot{\rho} = - i [\hat{H},\rho] + \sum_k \hat{L}_k \rho \hat{L}^\dag_k - \frac{1}{2}(\hat{L}^\dag_k \hat{L}_k\rho + \rho \hat{L}^\dag_k \hat{L}_k),
\end{equation}
as the quantum dots satisfy the Born-Markov approximation \cite{loss_quantum_1998,harbola_quantum_2006}.
The Hamiltonian $\hat{H}$ describes the unitary evolution while the Lindblad operators $\hat{L}_k$ capture the dissipative interaction with the environment.

Three levels of the quantum dots are used during gate operation. The coherent interactions of the system are described by
\begin{equation}
    \hat{H} = \delta \hat{a}^\dag \hat{a}+\hat{H}_{e,1}+\hat{H}_{e,2}+\hat{V}_1+\hat{V}_2.
\end{equation}
Here, $\hat{a}^{(\dagger)}$ are the annihilation (creation) operators of the cavity mode. The subscripts of the operators denote the qubit the respective coupling acts on:

The driving field $\hat{V}_j$ transfers population to the excited states with Rabi frequency $\Omega$,
\begin{equation}
    \hat{V}_j= \frac{\Omega}{2} (\ketbraind{e}{j}{0}+\ketbraind{0}{j}{e}). 
\end{equation}
The excited state Hamiltonian $\hat{H}_{e,j}$ contains the Jaynes-Cummings-type coupling that is essential to our engineering and the detuning of excited state $j$:
\begin{equation}
    \hat{H}_{e,j}= g(\hat{a}\ketbraind{e}{j}{0} + \hat{a}^\dag\ketbraind{0}{j}{e}) + \Delta \ketbraind{e}{j}{e}.
\end{equation}
The dissipative contribution we exploit to realize this nonunitary process is the spontaneous emission from the first qubit. It allows the excited state of the first qubit to decay into the state $\ket{1}$, represented by the following Lindblad operator:
\begin{equation}
    \hat{L}_{\gamma,1}=\sqrt{\gamma}\ketbraind{1}{1}{e}.
\end{equation}
The spontaneous emission also occurs on the second qubit with the same rate.
The cavity mode loses photons at rate $\kappa$.

There is a large amount of freedom in the choice of couplings. In Appendix\,\ref{ap:alternative} two alternative systems are introduced that perform the same gate action relying on oscillator decay as the dissipative contribution.

\section{Mechanism}\label{sec:Mechanism}

We use the couplings above in order to engineer the mapping $\ket{01}\rightarrow\ket{11}$ discussed in Sec.\,\ref{sec:Protocol}. 
This is achieved by tuning the parameters such that the desired decay becomes resonant while the undesired process is suppressed. 

\begin{figure}
    \centering
    \includegraphics[scale=0.5]{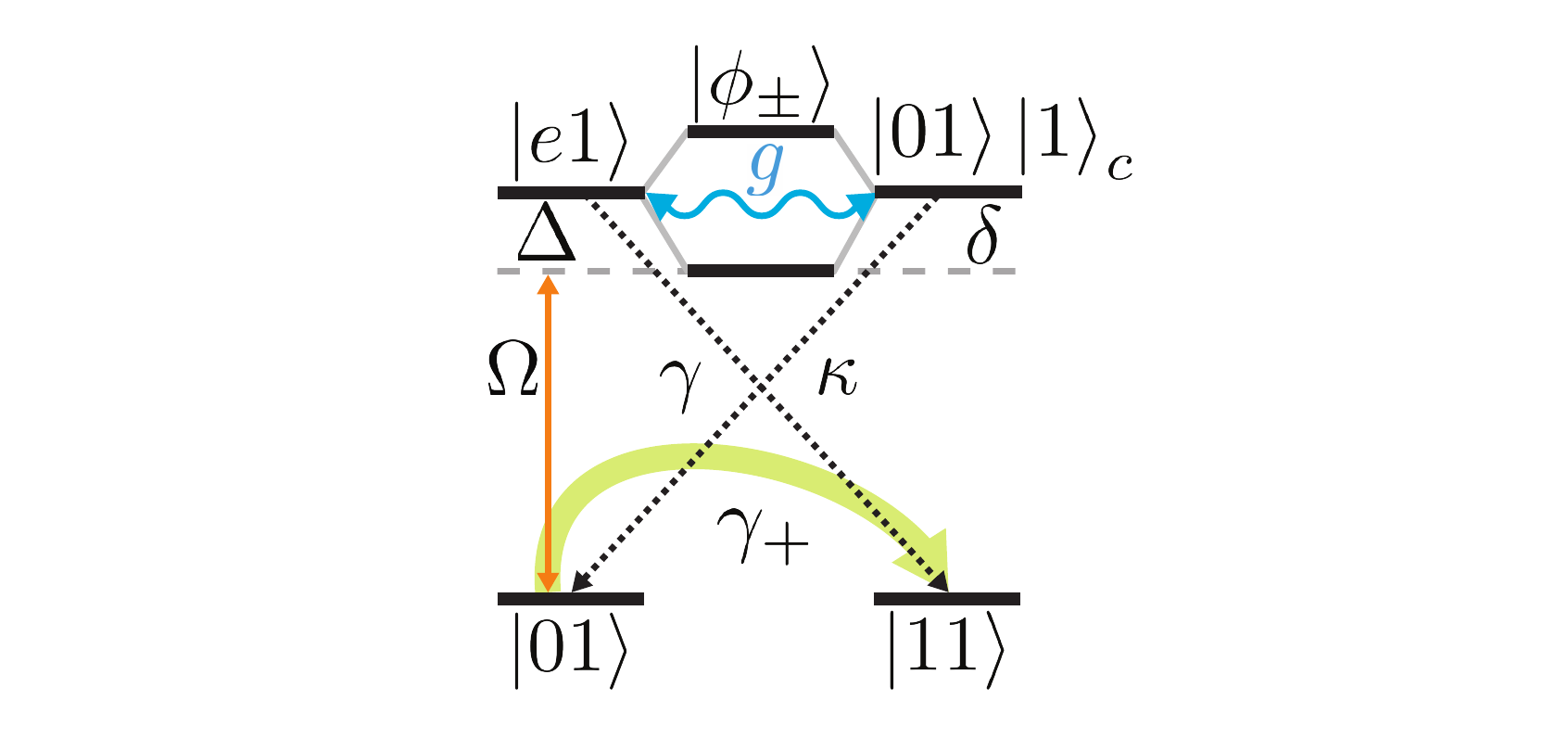}
    \caption{\textbf{Gate mechanism.} State $\ket{01}$ is off-resonantly driven to $\ket{e1}$ by the drive $\Omega$. The strong coupling $g$ results in the formation of dressed states $\ket{\phi_\pm}$, given in the text. For $\Delta\delta=g^2$ the lower dressed state $\ket{\phi_-}$ is in resonance with the drive and thus excited from $\ket{01}$. The gate mechanism is completed through subsequent spontaneous emission from $\ket{e1}$ to $\ket{11}$, realizing the desired action of an OR gate.}
    \label{fig:desired}
\end{figure}
The engineering is sketched in Fig.\,\ref{fig:desired}. Initial state $\ket{01}$ is coupled to $\ket{e1}$ by the weak drive $\hat{V}_1$. Due to the detuning $\Delta$ this drive would be off-resonant in absence of the Jaynes-Cummings-type coupling. For the choice $\delta = \Delta = g$ the strong coupling $g$ causes state $\ket{e1}$  to form dressed states 
\begin{equation}
    \ket{\phi_\pm} = \frac{1}{\sqrt{2}}\left(\ket{e1} \pm \ket{01}\ket{1}_c \right),
\end{equation}
with the cavity excited state $\ket{01}\ket{1}_c$.
For the chosen parameters, the lower energy dressed state $\ket{\phi_-}$ is in resonance with respect to the drive resulting in strong excitation. Subsequently, the state decays to the target state $\ket{11}$ through spontaneous emission.

\begin{figure}
    \centering
    \includegraphics[scale=0.5]{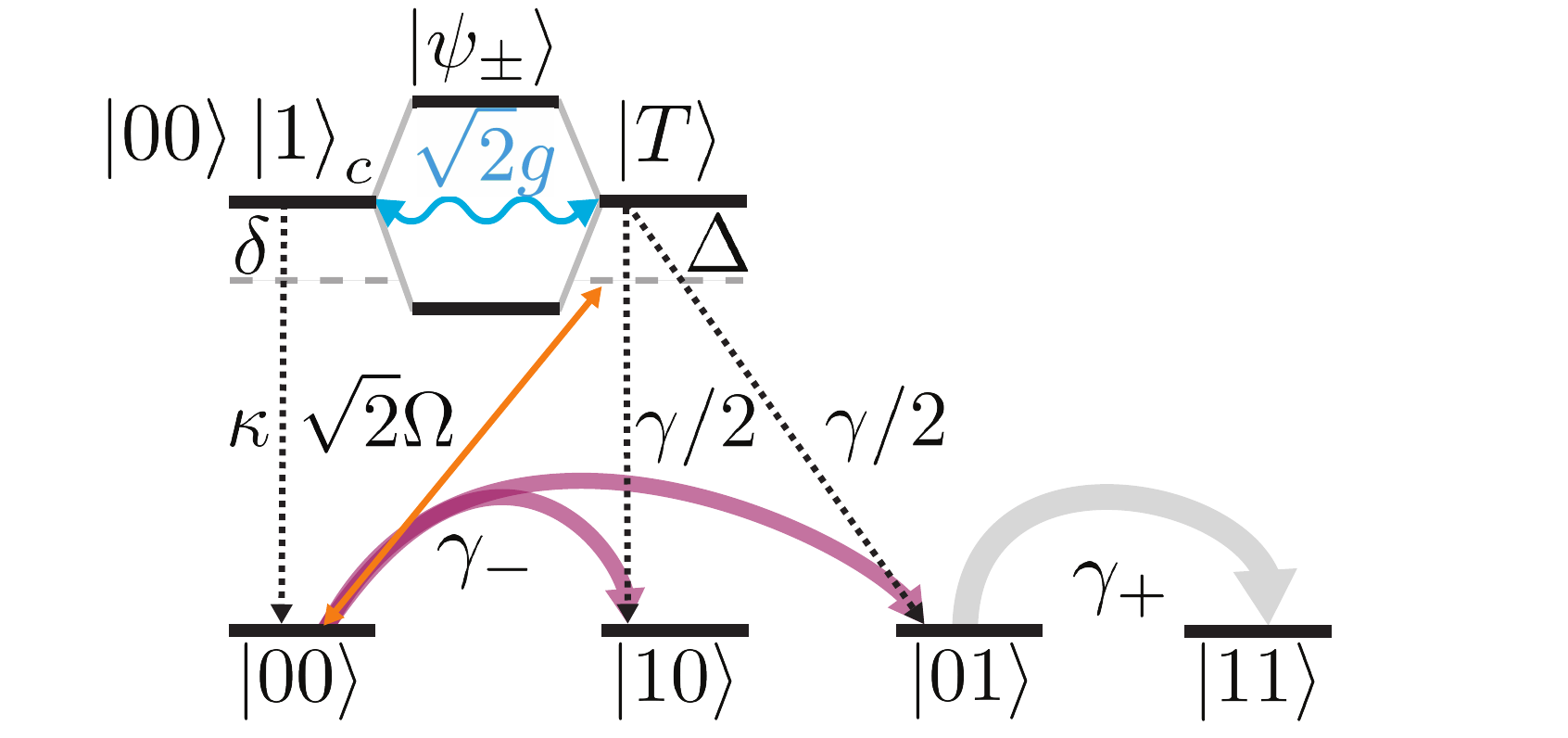}
    \caption{\textbf{Intrinsic error processes}. Ground state $\ket{00}$ is coherently driven to the excited state $\ket{T}=(\ket{0e}+\ket{e0})/\sqrt{2}$ which forms dressed states $\ket{\psi_\pm}$ with the cavity excited state $\ket{00}\ket{1}_c$. The dressed states $\ket{\psi_\pm}$ are given in the text. Due to constructive interference of the terms the coupling is enhanced to $\sqrt{2}g$. This shifts both dressed states $\ket{\psi_\pm}$ out of resonance with the drive leading to weak excitation of $\ket{00}$. The effective jump $\ket{00} \rightarrow \ket{10}$ producing an error is thus suppressed. The jump $\ket{00}\rightarrow\ket{01}$ does not produce an error, however the subsequent resonant process $\ket{01}\rightarrow \ket{11}$ with rate $\gamma_+$ does. }
    \label{fig:undesired}
\end{figure}

The gate action also entails error processes. Initial state $\ket{00}$ is driven to $\ket{T}=(\ket{0e}+\ket{e0})/\sqrt{2}$. It forms dressed states
\begin{equation}
    \ket{\psi_\pm}= \frac{1}{\sqrt{2}}\left(\ket{T} \pm \ket{00}\ket{1}_c \right),
\end{equation}
with the cavity excited state $\ket{00}\ket{1}_c$. Constructive interference between the terms in $\ket{\psi_\pm}$ increases the coupling strength to $\sqrt{2}g$. For the parameter choice $\delta = \Delta = g$ the dressed states have energies $\approx(1\pm\sqrt{2})g$ and are therefore both off-resonant with the drive. Consequently, the error process is much slower than the gate action (Fig.\,\ref{fig:undesired}).

 By adjusting $r=\frac{\Delta}{\delta}$ we can tune the coefficients of the dressed states. The choice $r = \frac{\gamma}{\kappa}$ maximizes the rate $\ket{01} \rightarrow \ket{11}$ (see Appendix\,\ref{ap:detunings}). For this choice the resonant dressed state shifts to
 \begin{equation}
    \ket{\phi_-} =\sqrt{\frac{\kappa}{\kappa+\gamma}}\ket{e1} - \sqrt{\frac{\gamma}{\kappa+\gamma}}\ket{01}\ket{1}_c.
\end{equation}

The effective operator formalism, detailed in \cite{reiter_effective_2012}, provides an alternative viewpoint of the gate action and allows us to analytically describe the system dynamics. It reduces the full dynamics of the system to the ground states. The mapping, $\ket{01}\rightarrow\ket{11}$ is implemented continuously by an effective jump operator that connects two ground states.

This effective description separates the system into ground and excited subspaces coupled by a weak drive $\hat{V}$.
The excited states are adiabatically eliminated and effective jump operators between the ground states remain. An effective decay process consists of a weak excitation with Rabi frequency $\Omega$, evolution within the excited subspace, and decay with the rate $\gamma$ or $\kappa$. 
The effective jump operators corresponding to spontaneous emission from the first and second qubits are:
\begin{align}
    \hat{L}^{\text{eff}}_{\gamma_1}&=\sqrt{\gamma_+}\ketbra{11}{01} +
   \sqrt{\gamma_-}\ketbra{10}{00}, \\
    \hat{L}^{\text{eff}}_{\gamma_2}&=\sqrt{\gamma_+}\ketbra{11}{10} +
   \sqrt{\gamma_-}\ketbra{01}{00}.
\end{align}
They are derived in Appendix \ref{ap:effopSpont}.
The rate of the desired process $\ket{01}\rightarrow\ket{11}$ is:
\begin{align}
    \gamma_+ =  \gamma \!\left(\!\frac{\Omega}{2}\!\right)^2 \!\abs{\frac{\tilde{\delta}}{\tilde{\delta}\tilde{\Delta}-g^2}}^2.
\end{align}
In the expression we use the complex detunings $\tilde{\Delta} = \Delta - \frac{i}{2}\gamma$ and $\tilde{\delta} = \delta -  \frac{i}{2}\kappa$.
The gate operation mediated by the desired decay process is shown in Fig.\,\ref{fig:desired}.
The undesired decay process $\ket{00}\rightarrow\ket{10}$ occurs with the rate:
\begin{align}
    \gamma_- =\gamma \!\left(\!\frac{\Omega}{2}\!\right)^2\! \abs{\frac{\tilde{\delta}}{\tilde{\delta}\tilde{\Delta}-2g^2}}^2.
    \label{eq:undesired}
\end{align}
It is shown in Fig.\,\ref{fig:undesired}.

We maximize the rate $\gamma_+$ by choosing
\begin{equation}
    \delta \Delta = g^2 ,\;\; \frac{\Delta}{\delta}=\frac{\gamma}{\kappa}.
    \label{eq:choice}
\end{equation}
The derivation of the effective rates and their optimization is performed in Appendix\,\ref{ap:effopSpont} and \ref{ap:detunings} respectively.
For strong coupling, $g^2 \gg \{\gamma \kappa, \kappa^2\}$, we obtain for the desired rate
\begin{align}\label{eq:gammaplus}
    \gamma_+^\text{opt} \approx \frac{\Omega^2}{4 \gamma}.
\end{align}
The rate for the undesired rate in Eq. \eqref{eq:undesired} is, in turn,
\begin{align}\label{eq:gammaminus}
    \gamma_-^\text{opt} \approx \frac{\gamma \Omega^2}{4 g^2}.
\end{align}
Physically, the parameter choice in Eq. \eqref{eq:choice} shifts the lower energy dressed state $\ket{\phi_-}$ into resonance with the driving field $\hat{V}$. 
Consequently, we have that $\gamma_+ \gg \gamma_-$. This ensures that the desired process occurs at a much higher rate than the undesired process. The resulting operators are then well approximated by
\begin{align}
    \hat{L}^{\text{eff}}_{\gamma_1}
    &\approx \sqrt{\gamma_+}\ketbra{11}{01},
\\
    \hat{L}^{\text{eff}}_{\gamma_2} 
    &\approx\sqrt{\gamma_+}\ketbra{11}{10}.
\end{align}
We thus realize the desired action of the OR gate, $\ket{01} \rightarrow \ket{11}$. As a byproduct, the effective decay $\ket{10} \rightarrow \ket{11}$ is also on resonance and enhanced; however, this does not create errors.

Photon loss in the cavity causes dephasing described by the following effective Lindblad operator
\begin{align}
\hat{L}^\text{eff}_{\kappa} = \sqrt{\kappa}&\frac{\Omega}{2} \biggl[ \frac{2}{ g_\text{eff,3}}\ketbra{00}{00} 
+ \frac{1}{ g_\text{eff,2}}\ketbra{10}{10}\nonumber\\
&+\frac{1}{ g_\text{eff,2}}\ketbra{01}{01}   \biggr].
\end{align}
Here $g_{\text{eff},2}= g-\frac{\tilde{\delta}\tilde{\Delta}}{g}$ and $g_{\text{eff},3} = 2g - \frac{\tilde{\delta}\tilde{\Delta}}{g}$. The dephasing does not present an issue to our gate as we aim for a classical operation.

An analytic description of the dynamics of our gate using the effective operators and other related methods is presented in Appendix\,\ref{ap:AnalyticEvo}.

\section{Performance}\label{sec:Performance}
Every physical gate operation is subject to errors. The quantum gate fidelity is a tool to measure a gate's accuracy. When aiming to implement the operation $\mathcal{G}$ we might instead implement the noisy operation $\mathcal{E}.$ The \textit{gate fidelity} measures how close $\mathcal{G}$ and $\mathcal{E}$ are given input state $\rho$ and is defined as $\mathcal{F}(\mathcal{E}(\rho) , \mathcal{G}(\rho))$. 
For our classical gates, only the state of the output qubit is relevant. Therefore, before we measure the fidelity we first trace over all qubits except the output. We call this measure the \textit{success probability}. For input state $\ket{01}$ in the OR gate the success probability
\begin{equation}
    P_s(01)=\mathcal{F}(\Tr_2[{\mathcal{E}(\ketbra{01}{01})}],\ketbra{1}{1}),
\end{equation}
measures the probability to be in an a state with $\ket{1}$ on the first qubit after the operation has been completed.  The \textit{error probability} is defined as 
\begin{equation}
P_e(x) = 1-P_s(x).
\end{equation}
To remove the dependence of the input state we define the \textit{average error} as
\begin{equation}
P_e^{avg}= \average_{x\in S} P_e(x).
\end{equation}
Here $S$ is the set of computational basis states.
\subsection{Numerical analysis}
We assess the performance of the gate using numerical simulation. To this end we perform a simulation in QuTIP, an open-source python library for simulating quantum systems \cite{johansson_qutip_2013}. The infinite Fock space of the harmonic oscillator was truncated after two excitations.
To investigate error rates attainable by a quantum dot setup we assume experimentally realistic parameters $(\frac{g}{2 \pi})=4.4$ GHz, $(\frac{\gamma}{2 \pi})=0.3$ GHz, $(\frac{\kappa}{2 \pi} )= 0.6$ GHz \cite{najer_gated_2019}. Additionally we consider noise processes on the logical states which happen at a lower rate with $ T_1 = 20   \mu $s, $ T_2 = 1 \mu$s. The incoherent flips of the ground states are captured by
\begin{equation}
    \hat{L}_{+\gamma_g,j}= \sqrt{\gamma_g} \ketbraind{1}{j}{0}, \quad \hat{L}_{-\gamma_g,j}= \sqrt{\gamma_g} \ketbraind{0}{j}{1}.
\end{equation}
They occur symmetrically on both qubits.

\begin{figure*}
\includegraphics{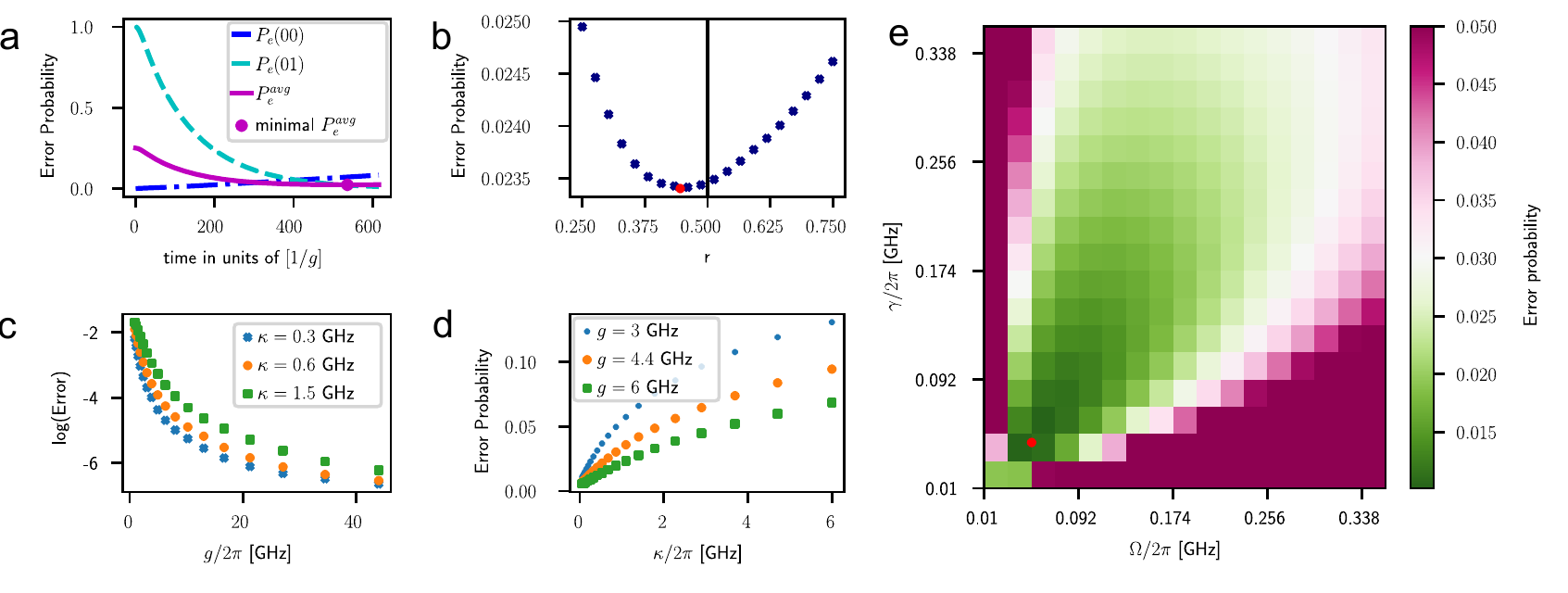}
\caption{\label{fig:wide}\textbf{Performance} (a) How the error probability is evaluated. For a set of parameters, the time evolution of all classical input states is simulated. From this, we calculate the average error probability. By minimizing this over time the optimal gate time and minimal gate error are found. Initially, $\ket{01}$ has a high error probability as it has the wrong value on the output qubit. It is quickly mapped to $\ket{11}$ by rate $\gamma_+$. The error probability of state $\ket{00}$ increases over time due to the undesired decay $\gamma_-$ and ground state errors $\gamma_g$. (b) Balance of the detunings. The product of the cavity and excited state energies is fixed by the resonance condition $g^2=\delta\Delta$. By adjusting $\Delta = \sqrt{r} g$ and $\delta = g/\sqrt{r}$ with the parameter $r$ we can tune the coefficients of $\ket{e1}$ and $\ket{01}\ket{1}_c$ within the dressed states. Analytically the optimum is found to be at $r=\gamma/\kappa$ which evaluates to $r=1/2$ for the experimental parameters. The simulated optimum closely matches the analytic optimum. (c) Change of $g$. A higher $g$ shifts the dressed states $\ket{\psi_\pm}$ further out of resonance, thereby reducing the rate of the undesired decay. (d)   Change of $\kappa$. The oscillator decay with rate $\kappa$ does not contribute to the action of the gate. Independent of the value of $g$ a low $\kappa$ improves the performance. (e) Two-dimensional optimization of $\gamma$ and $\Omega$. The remaining parameters are fixed to experimentally feasible parameters $(\frac{g}{2 \pi})=4.4$ GHz, $(\frac{\kappa}{2 \pi} )= 0.6$ GHz. $\Delta$ and $\delta$ are fixed by the condition discussed in Panel (b). At $(\frac{\Omega}{2\pi}) = 0.04$ GHz, $(\frac{\gamma}{2\pi}) = 0.03$ GHz a minimal error probability 0.01 is reached. The operation is robust to parameter choice; for a wide range of system parameters and also gate times the error is close to minimal.}
\end{figure*}
The error of the gate is limited by the undesired effective decay and decoherence within the ground states. Dephasing does not diminish the performance, even though it occurs at a much larger rate, as the operation is inherently classical.
We optimize the parameters by minimizing $P_e^\text{avg}$. We have to tune six parameters $\delta, \Delta, g, \gamma, \kappa,$ and $\Omega$.  The resonance condition for the desired process $\delta \Delta = g^2$, fixes one of them.

Figure\,\ref{fig:wide} (a) shows how the minimal error probability for a given set of parameters is calculated. We simulate the time evolution of all possible initial states and evaluate the maximal error in every time step. Initial state $\ket{01}$ has to be mapped to state $\ket{11}$, therefore it has a high error probability at first. The state $\ket{00}$ should not be mapped anywhere; however, due to the undesired decay $\gamma_-$ and the ground state flips its error probability increases over time. After the optimal gate time the average error probability, $P_e^{avg}$, is minimal. This minimal error is the gate error probability of this set of parameters. 

We can tailor the coefficients of $\ket{e1}$ and $\ket{01}\ket{1}_c$ in the dressed state $\ket{\phi_-}$ by adjusting the ratio of the detunings $r=\frac{\Delta}{\delta}$. Thereby, we adjust the width and the Rabi frequency addressing the dressed state. The analytic optimum $r= \frac{\gamma}{\kappa}$ is found by maximizing the effective decay rate $\gamma_+$ in Appendix\,\ref{ap:detunings}. If we assume the experimentally feasible parameters and only adjust the driving strength $\Omega$ and the detunings $\delta$ and $\Delta$ the minimal error of $2.3 \%$ is attained at $r =0.44$ and $(\frac{\Omega}{2\pi}) = 0.13$ GHz (Fig.\,\ref{fig:wide} (b)). The discrepancy between the analytic and numeric optimum arises because the analytic optimization does not take the undesired process into account. 

The influence of the cavity decay rate $\kappa$ on the performance can be seen in Fig.\,\ref{fig:wide} (c).
The cavity decay does not contribute to the action of the gate. A higher $\kappa$ leads to level broadening increasing the rate of the undesired decay $\gamma_-$ relative to the desired decay $\gamma_+$. 

Improving the coupling to the cavity $g$ only reduces the error. A higher $g$ leads to more pronounced splitting of the dressed states thereby reducing excitations of state $\ket{00}$. Additionally, with an increase in $g$ all couplings within the excited state Hamiltonian and the drive can also be increased. This accelerates the gate action relative to the flips in the ground states, reducing the error probability (Fig.\,\ref{fig:wide} (d)). 

Spontaneous emission contributes both to the undesired decay and the action of the gate. The necessary and undesired effects of spontaneous emission give rise to an optimum. A too high value of $\gamma$ leads to level broadening and allows the undesired decay to occur. A too low value of $\gamma$ makes the gate action slow and allows ground state errors to occur. 
Similarly, the drive $\Omega$ also has an optimum. If $\Omega$ is too low the gate action is slow and ground state errors occur. Increasing $\Omega$ does not equally increase $\gamma_+$ and $\gamma_-$ as saturation effects slow the resonant decay $\gamma_+$ \cite{reiter_scalable_2016}. A value $\Omega$ larger than $\gamma$ is therefore also undesirable. These effects are shown in Fig.\,\ref{fig:wide} (e). The optimum combination is found to be $(\frac{\Omega}{2\pi}) = 0.10$ GHz, $(\frac{\gamma}{2\pi}) = 0.08$ GHz at a minimal error probability $1 \%$.

\subsection{Energy consumed by the operation}

Irreversible classical computation is linked to dissipation by Landauer's principle \cite{landauer_irreversibility_1961}. This principle is a lower bounds the energetic cost of information erasure. In Appendix\,\ref{ap:Scattered} we estimate the number of absorbed photons during operation. This serves as a first estimate of the energetic cost of our operation. By expanding on this consideration one could put our gates into the context of Landauer's principle

\section{Further gate operations}
\label{sec:FurtherGates}

The protocol used to create the OR gate can also be applied to realize other classical gates. The AND gate directly follows from the OR as it is structurally equivalent. If we relabel $\ket{0}\leftrightarrow\ket{1}$ on both qubits the OR produces an AND gate. Either gate together with a NOT gate is functionally complete \cite{wernick_complete_1942}. Further gates such as the XOR gate and the functionally complete NOR gate can be realized by splitting the evolution into separate steps for excitation and decay. Possible protocols and gate mechanisms are discussed in Appendix \ref{ap:XOR} and \ref{ap:NOR} respectively.

\section{Conclusion}\label{sec:Conclusion}
We have presented a novel way of performing classical logic gates in quantum systems. By utilizing dissipation, we engineered the required nonunitary evolution on the smallest stable Hilbert space. Our approach works without the need for measurement or ancilla qubits. 

Furthermore, we demonstrate that the OR gate can be implemented in realistic quantum dot setups. Two quantum dots in a Voigt magnetic field have a suitable level structure and the necessary decay processes to create the required mapping. We utilize spontaneous emission as the dissipative contribution. Using numerical and analytical tools, we optimized the system parameters in the presence of noise. In doing so we reached a minimal average error below $2.5 \%$.

The presented design principles enrich the toolbox of quantum engineering.
It would be of interest to investigate their application to further irreversible tasks in quantum information processing. In cases where operations conditioned on measurements are required, our operations may be applicable, reducing the need for classical measurement and feedback/feedforward.
One possible application may be given in measurement-based quantum computing. In this universal paradigm for quantum computation sequences of adaptive measurements are performed on an entangled cluster state \cite{gottesman_demonstrating_1999,briegel_measurement-based_2009}. These adaptive measurements require classical computation.
A second possible application may be found in quantum error correction. The correction operations are performed conditioned on the result of syndrome measurements \cite{nielsen_quantum_2010,reiter_dissipative_2017}. %
Similarly, quantum convolutional neural networks use unitaries conditioned on measurements to mimic the effect of pooling layers in classical neural networks \cite{cong_quantum_2019,herrmann_realizing_2021}.
Here our gates could be used to allow for more general interactions between measured qubits.
While currently these operations are performed by measuring individual qubits, generalizations of the two-qubit operations introduced in this paper could be used to realize such tasks. This may allow reducing the number of ancillae, classical measurements, and feedforward operations needed during operations. Such improvements are crucial in the NISQ era, where the qubit count is limited. Furthermore, by reducing the number of qubits also the circuit depth may be reduced.

\section{Acknowledgments}
The authors thank Tabea Bühler, Ivan Rojkov, Jonathan Home, and Isaac L. Chuang for their questions and valuable discussions. We acknowledge funding from the Swiss National Science Foundation (Ambizione grant no. PZ00P2 186040) and the ETH Research Grant ETH-49 20-2. A. J. acknowledges support from the European Union's Horizon 2020 Research and Innovation Programme under Marie Skłodowska-Curie grant agreement no. 840453 (HiFig).

\bibliography{references}

\appendix

\section{Alternative implementations}\label{ap:alternative}
The couplings that realize the OR-gate are not unique. Various sources of dissipation and excited state couplings can realize the mappings needed for the OR gate. By considering different couplings and other sources of dissipation we remain flexible regarding the choice of the physical system and facilitate future experimental implementations.
\subsection{Oscillator-decay-based scheme}
Oscillator decay can also be utilized to perform the action of the gate.
The Hamiltonian for this system is
\begin{equation}
     \hat{H} = \delta \hat{a}^\dag \hat{a}+\hat{H}_{e,1}+\hat{H}_{e,2}+\hat{V}_1.
\end{equation}{}
We adjust the excited state Hamiltonian
\begin{equation}
    \hat{H}_{e,j}=g(\hat{a}\ketbraind{e}{j}{1} + \hat{a}^\dag\ketbraind{1}{j}{e}) + \Delta \ketbraind{e}{j}{e}  ,
\end{equation}{}
to couple $\ket{1}\leftrightarrow\ket{e}$. 
The decay of the oscillator 
\begin{equation}
    \hat{L}_c=\sqrt{\kappa} \hat{a}, \; \kappa=\gamma,
\end{equation}
completes the action of the gate. The desired decay now happens over an excited subspace with three states; therefore, the resonance condition is $2g^2=\delta\Delta$. The effective operators of these couplings are derived in Appendix\,\ref{ap:OsziEffop}.

\subsection{Hybrid scheme}
A hybrid between the spontaneous emission and oscillator decay based schemes is achieved through the couplings
\begin{equation}
    \hat{H}_{e,1}=\Delta \ketbraind{e}{1}{e} + g(\hat{a}\ketbraind{e}{1}{1} + \hat{a}^\dag{}\ketbraind{1}{1}{e}),
\end{equation}{}
\begin{align}
\hat{H}_{e,2} =  \Delta \ketbraind{e}{2}{e} + g( \hat{a}\ketbraind{e}{2}{0} + \hat{a}^\dag{}\ketbraind{0}{2}{e}).
\end{align}
One should note that the Jaynes-Cummings-type coupling is asymmetric i.e we couple $\ket{e}_1 \leftrightarrow \ket{1}_1$ on the first qubit and $\ket{e}_2 \leftrightarrow \ket{0}_2$ on the second. The coherent driving field $\hat{V}$ needs to individually address the first qubit as also driving the second field would introduce resonant error processes
\begin{equation}
    \hat{V}_1= \frac{\Omega}{2} (\ketbraind{e}{1}{0}+\ketbraind{0}{1}{e}).
\end{equation}{}
In this system, the oscillator contributes to the gate action similar to the oscillator-based scheme. The desired decay happens over a two-state excited subspace just as in the first scheme. Therefore, the resonance condition is the same, $g^2=\delta\Delta$.

\section{NOR gate}\label{ap:NOR}
 The NOR gate is a highly desirable two-qubit gate as it alone is functionally complete, provided one can initialize qubits in any logical state and can copy inputs \citep{enderton_mathematical_2001}. 
 
 \begin{figure}
     \centering
     \includegraphics[scale = 0.5]{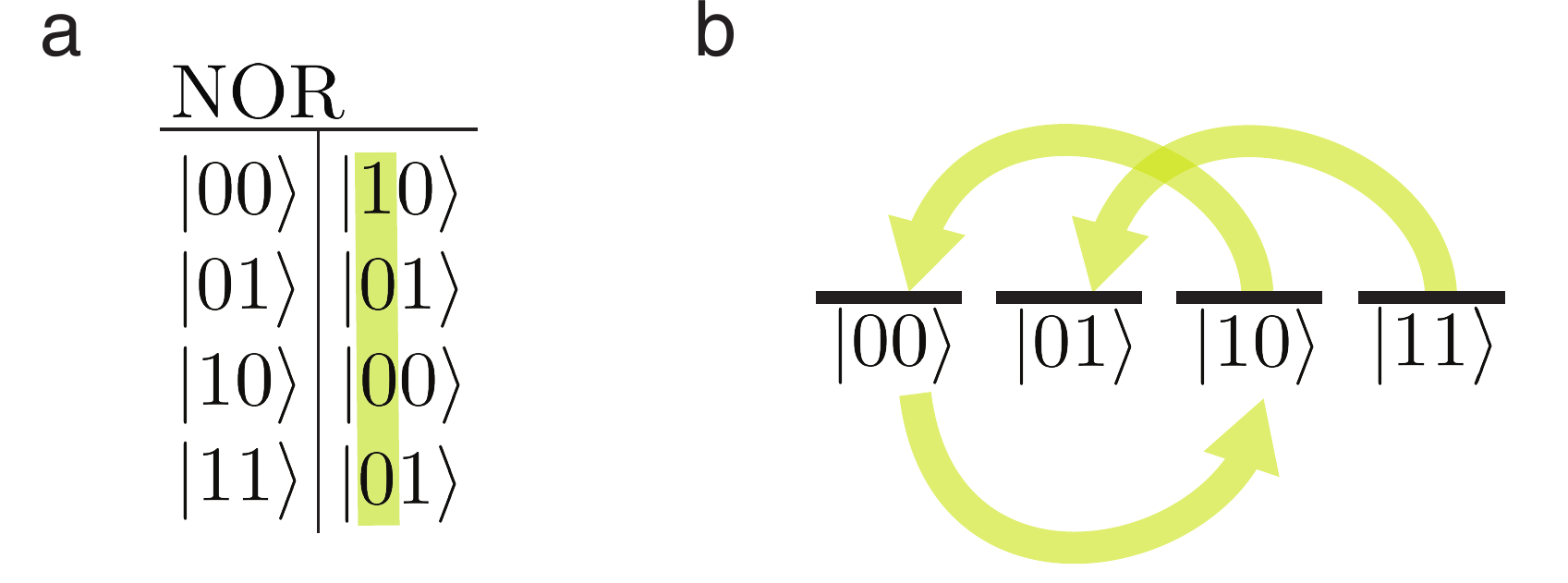}
     \caption{\textbf{NOR mappings} (a) The truth table of the NOR gate. The gate action is performed by swapping $\ket{00}$ and $\ket{10}$ and mapping $\ket{11}$ to $\ket{01}$. (b) The mappings we need to engineer in a level scheme.}
     \label{fig:NOR_idea}
 \end{figure}
In Fig.\,\ref{fig:NOR_idea} the necessary mappings for a NOR gate are illustrated.
If we were to apply these mappings continuously both inputs $\ket{00}$ and $\ket{10}$ would result in a mixed output. This mixing occurs due to multiple decays between state $\ket{00}$ and $\ket{10}$. 
Any mapping that satisfies the truth table of the NOR gate suffers from the same or related issues. We could try to get around this problem by displaying the output on the second instead of the first qubit. This does not produce the desired effect as the NOR gate is symmetric, (i.e $\ket{01}$ and $\ket{10}$ are mapped to the same output) as a consequence, changing the output qubit leaves the structure of mappings the same. 

We can bypass the issue by separating the evolution into two distinct steps. First, the ground states are transferred to the excited states through a $\pi/2$ pulse. Only after this excitation has been completed the excited states are allowed to decay. This separates the evolution of initial states $\ket{00}$ and $\ket{10}$ allowing us to implement the mappings shown in Fig.\,\ref{fig:NOR_idea}. Now all computational basis states evolve separately therefore their output is not mixed.

In addition to the couplings used for the OR gate further resources are required. The mappings $\ket{10}\rightarrow\ket{00}$ and $\ket{11}\rightarrow\ket{01}$  can be achieved by optical pumping. If the first qubit is set to $\ket{1}$ it should be flipped, independent of the state of the second qubit. 
To map $\ket{00}\rightarrow\ket{10}$ we have to make use of resonance engineering just as we did for the OR gate. The mapping corresponds to the undesired decay of the OR gate. This allows us to use any of the systems we used for the OR gate with a different resonance condition.

\subsection{System}
We consider two four-level systems coupled to a harmonic oscillator. The systems consist of two logical ground states ($\ket{0}$ and $\ket{1}$) and two excited states ($\ket{e}$ and $\ket{f}$) used during gate operation. State $\ket{e}$ strongly couples to a harmonic oscillator to enable resonance engineering. 
We separate the system Hamiltonian into the excited state Hamiltonians and the driving fields
\begin{equation}
    \hat{H} = \delta \hat{a}^\dag \hat{a}+\hat{H}_\text{e,1}+\hat{H}_\text{e,2}+\hat{V}_\text{e,1}+\hat{V}_\text{f,2}.
\end{equation}
The excited states evolve according to:
\begin{equation}
    \hat{H}_{e,j}=\Delta \ketbraind{e}{j}{e}+ g(\hat{a}\ketbraind{e}{j}{1} + \hat{a}^\dag\ketbraind{1}{j}{e}).
\end{equation}{}
In addition to the drive on $\ket{0}$
\begin{equation}
    \hat{V}_e= \frac{\Omega_e}{2}(\ketbraind{e}{1}{0}+\ketbraind{0}{1}{e}),
\end{equation}
we also drive state $\ket{1}$ to $\ket{f}$
\begin{equation}
    \hat{V}_f= \frac{\Omega_f}{2}(\ketbraind{f}{1}{1}+\ketbraind{1}{1}{f}).
\end{equation}
Oscillator-decay is the dissipative resource for the process $\ket{00}\rightarrow\ket{10}$ created by engineered dissipation. It is described by the Lindblad operator:
\begin{equation}
    \hat{L}_\kappa = \sqrt{\kappa}\hat{a} .
\end{equation}
The optical pumping uses spontaneous emission
\begin{equation}
    \hat{L}_\gamma = \sqrt{\gamma}\ketbraind{0}{1}{f}.
\end{equation}
For the stepwise implementation, these decays must not be present during the excitation process. In trapped ions, one could create the rate $\gamma$ by repumping state $\ket{f}$ to some auxiliary level that decays rapidly. To create the oscillator-decay one would use sympathetic cooling.

\subsection{Mechanism}

\begin{figure}
    \centering
    \includegraphics{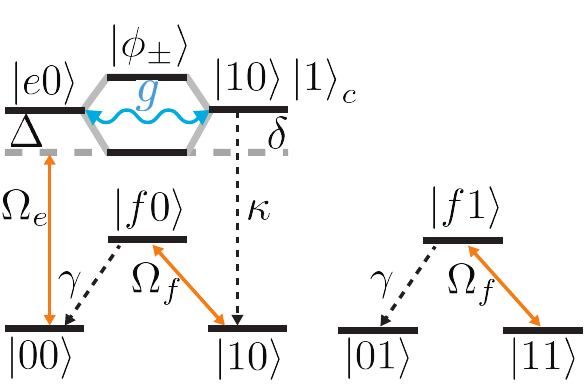}
    \caption{\textbf{NOR desired processes.} The decay $\ket{00}\rightarrow\ket{10}$ could be realized by engineered dissipation using the same couplings as the oscillator-decay-based scheme. The resonance condition changes to $\delta\Delta=g^2$ such that the drive $\ket{00}\leftrightarrow \ket{\phi_-}$ is resonant. The mappings $\ket{10} \rightarrow \ket{00}$ and $\ket{11} \rightarrow \ket{01}$ do not require any selectivity. The state transfer is carried our by optical pumping.}
    \label{fig:NOR_desired}
\end{figure}
The gate operates through the mappings illustrated in Fig.\,\ref{fig:NOR_idea}. 
The desired mapping $\ket{00}\rightarrow\ket{10}$ is shown together with the optical pumping as a level scheme in Fig.\,\ref{fig:NOR_desired}. Initial state $\ket{00}$ is off-resonantly driven to $\ket{e0}$. The strong interaction within the excited-states couples $\ket{e0}$ to $\ket{10}\ket{1}_c$. They form dressed states $\ket{\phi_\pm}$. The lower energy dressed state $\ket{\phi_-}$ is shifted into resonance with the drive for $\delta\Delta=g^2$. After the excitation is completed the oscillator excited-state is allowed to decay through sympathetic cooling.
Simultaneously the drive with Rabi frequency $\Omega_f$ resonantly addresses the $\ket{1}_1\leftrightarrow\ket{f}$ transition. The $\pi/2$ pulse transfers the entire population of $\ket{10}$ and $\ket{11}$ to the excited-state $\ket{f0}$ and $\ket{f1}$. When all excitations have been completed state $\ket{f}$ is allowed to decay to $\ket{0}$.
\begin{figure}
    \centering
    \includegraphics{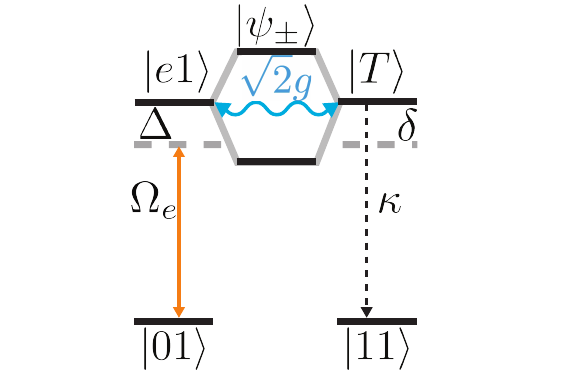}
    \caption{\textbf{NOR undesired processes.} The engineered dissipation also gives rise to an error process. Ground-state $\ket{01}$ is driven to the excited-state $\ket{e1}$. Together with the state $\ket{\text{T}}=(\ket{11}\ket{1}_c+\ket{1e})/\sqrt{2}$ it forms dressed states $\ket{\psi_\pm}$. They couple strongly with $\sqrt{2}g$ shifting them out of resonance with the drive. }
    \label{fig:NOR_undesired}
\end{figure}
The undesired process, shown in Fig.\,\ref{fig:NOR_undesired}, creates errors. Initial state $\ket{01}$ is off-resonantly to $\ket{e1}$. It couples to $\ket{11}\ket{1}_c$ and $\ket{1e}$. Due to the constructive interference between the terms the coupling increases to $\sqrt{2}g$. This shifts both dressed state $\ket{\psi_\pm}$ out of resonance resulting in weak excitation.

We can use the effective operator formalism to better understand the dynamics of the gate. However, we have to keep in mind that not all couplings will be present simultaneously. The effective jump operator for oscillator decay is:
\begin{align}
    \hat{L}_{\kappa}^\text{eff} =- \sqrt{\kappa} \frac{\Omega_e}{2}\left[ \frac{1}{g_{\text{eff},3}}\ketbra{11}{01} + \frac{1}{g_{\text{eff},2}}\ketbra{10}{00}\right],
\end{align}
with the effective couplings $g_{\text{eff},3}= 2g-\frac{\tilde{\delta}\Delta}{g}$ and $g_{\text{eff},2}=g-\frac{\tilde{\delta}\Delta}{g}$.
For the map $\ket{00}\rightarrow\ket{10}$ to be resonant we choose $\delta\Delta = g^2$. The resulting effective jump operator is:
\begin{align}
    \hat{L}_{\kappa}^\text{eff} = \sqrt{\gamma_-}\ketbra{11}{01} +\sqrt{\gamma_+}\ketbra{10}{00}
\end{align}
Here we used the effective rate $\gamma_- = \frac{\Omega^2\kappa}{4g^2}$ and $\gamma_+ = \frac{\Omega^2}{4\kappa}$. Using $\kappa \ll g$ we can approximate
\begin{align}\label{eq:NorEngineered}
    \hat{L}_{\kappa}^\text{eff}\approx \sqrt{\gamma_-} \ketbra{11}{01}.
\end{align}
For the optical pumping we obtain the effective jump operator:
\begin{align}\label{eq:optpump}
    \hat{L}_\gamma^\text{eff} = \sqrt{\frac{\Omega_f^2}{\gamma}} \left( \ketbra{01}{11} +\ketbra{00}{10}\right).
\end{align}
It describes the mapping $\ket{11}\rightarrow\ket{01}$ and $\ket{10}\rightarrow\ket{00}$.

Thus the couplings realize the action of the NOR gate. For the engineered decay the desired process occurs at a far higher rate than the undesired process, see Eq. \ref{eq:NorEngineered}. The optical pumping, Eq. \ref{eq:optpump}, operates just as optical pumping for state preparation. The principle of operation is remains unchanged by operating in stepwise fashion \cite{lin_dissipative_2013}. 
\section{XOR gate}\label{ap:XOR}
The XOR gate is an irreversible classical logic gate. It maps inputs with even parity to 0 and odd parity inputs to 1. Its irreversibility arises only due to the reduction of output size.  If we retain the second qubit in the result the operation can be performed reversibly. 
The commonly used CNOT gate realizes an XOR gate with the output displayed on the second qubit. 

\begin{figure}
    \centering
    \includegraphics{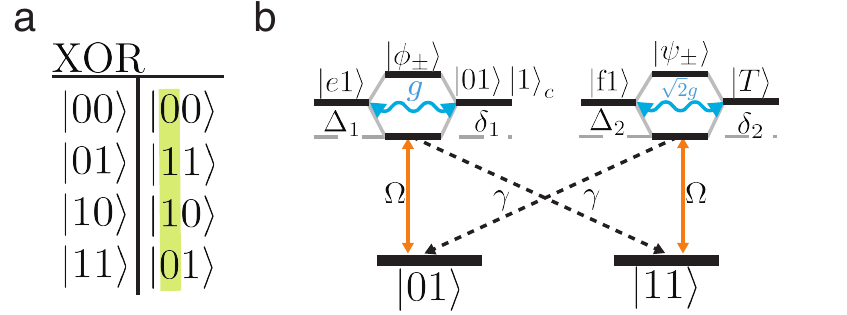}
    \caption{\textbf{Stepwise XOR.} (a) Truth table. To realize a XOR gate with the result on the first qubit we have to swap $\ket{01}\leftrightarrow \ket{11}$. (b) Possible gate operation. The excitation and decay have to occur in separate steps. The mapping $\ket{01}\rightarrow\ket{11}$ can be realized using any of the schemes discussed for the OR gate. The mapping $\ket{11}\rightarrow\ket{01}$ requires an additional $\ket{f}$ excited-state and harmonic oscillator mode $\hat{b}$.}
    \label{fig:XOR_stepwise}
\end{figure}

A stepwise approach can also be used to construct an XOR gate. If we display the result on the first qubit we can operate the gate by swapping $\ket{01}\leftrightarrow\ket{11}$. This swap can be achieved by two dissipative mappings operating in stepwise fashion.
The two mappings can be constructed analogously to the OR gate discussed in this paper. The truth table and a possible scheme are shown in Fig.\,\ref{fig:XOR_stepwise}. The operation is inspired by the spontaneous emission-based scheme. We assume a system with two distinct oscillator modes $\hat{a}$ and $\hat{b}$. In addition we require two exited states $\ket{e}$ and $\ket{f}$. The Hamiltonian describing the system is
\begin{align}
    \hat{H} = &\delta_1 \hat{a}^\dag \hat{a} +\hat{H}_{e,1}+\hat{H}_{e,2}+\hat{V}_{e,1} \nonumber \\
    &+\delta_2 \hat{b}^\dag \hat{b}+\hat{H}_{f,1}+\hat{H}_{f,2}+\hat{V}_{f,1},
\end{align}{}
In essence it doubles the excited degrees of freedom of the OR gate. We now consider two excited state Hamiltonians
\begin{align}
    \hat{H}_{e,j}=& \Delta_1 \ketbraind{e}{j}{e} 
    + g(\hat{a}\ketbraind{e}{j}{0} +\hat{a}^\dag{}\ketbraind{0}{j}{e} ),\\
    \hat{H}_{f,j}=&\Delta_2 \ketbraind{f}{j}{f} + g(\hat{b}\ketbraind{f}{j}{1} +\hat{b}^\dag{}\ketbraind{1}{j}{f} ).
\end{align}{}
We also add a second coherent fields
\begin{align}
    \hat{V}_{e,1}=& \frac{\Omega}{2}(\ketbra{e}{0}_1+\ketbra{0}{e}_1),\\
    \hat{V}_{f,1}=& \frac{\Omega}{2}(\ketbra{f}{1}_1+\ketbra{1}{f}_1).
\end{align}{}
The dissipative contributions we utilize are spontaneous emission from the excited-states of the first qubit
\begin{equation}
    \hat{L}_e = \sqrt{\gamma} \ketbra{1}{e}_1, \;\; \hat{L}_f = \sqrt{\gamma} \ketbra{1}{f}_1. 
\end{equation}

The calculation of the effective jump operators is identical to that in Appendix\,\ref{ap:effopSpont}. The couplings create the following effective jump operators:
\begin{align}
           \hat{L}^{\text{eff}}_{e}=\sqrt{\gamma }\frac{\Omega}{2} \biggl[\frac{1}{\Delta_\text{eff,1}^{1}}\ketbra{11}{01} +\frac{1}{\Delta_\text{eff,2}^{1}} \ketbra{10}{00} \biggl],
\end{align}
\begin{align}
    \hat{L}^{\text{eff}}_{f}=\sqrt{\gamma }\frac{\Omega}{2} \biggl[
    \frac{1}{\Delta_\text{eff,1}^{2}}\ketbra{00}{10}+\frac{1}{\Delta_\text{eff,2}^{2}} \ketbra{01}{11} \biggl].
\end{align}
Here we used the effective detunings
\begin{align}
    \Delta_\text{eff,1}^{j} &= \tilde{\Delta}_{j} -\frac{g^2}{\delta_{j}},\\
    \Delta_\text{eff,2}^{j} &= \tilde{\Delta}_{j} - \frac{g^2}{\delta_{j}-g^2/\delta_{j}}.
\end{align}
$\hat{L}_{\text{eff}}^{e}$ is the exactly the same effective jump operator as in the spontaneous emission-based scheme. We want the $\ket{01}\rightarrow\ket{11}$ term to be resonant, therefore we choose $\delta_1\Delta_1 = g^2$. Due to the similar couplings containing $\ket{f}$ and $\ket{e}$ only the labels in the jump operator $\hat{L}_{\text{eff}}^{f}$ change. We want the decay $\ket{11}\rightarrow\ket{01}$ to be resonant. Therefore, we choose $\Delta_2\delta_2 = 2 g^2$.
The analysis of the contrast between undesired and desired rates mirrors that of the OR gate (see Sec.\,\ref{sec:Mechanism}). Thus we conclude that our choice of couplings realizes the XOR gate.
\section{Effective system dynamics}
Here we employ the \textit{effective operator formalism} to obtain the effective dynamics of the ground states \cite{reiter_effective_2012}. 
The dynamics are governed by effective Hamiltonian and Lindblad operators
\begin{align}
    \hat{H}_\text{eff} = - \frac{1}{2}& \hat{V}_- [ \hat{H}_\text{NH}^{-1} + (\hat{H}_\text{NH}^{-1})^\dag ] \hat{V}_+ + \hat{H}_g ,\label{eq:effhamiltonian} \\
    \hat{L}_\text{eff}^k& = \hat{L}^k \hat{H}_\text{NH}^{-1} \hat{V}_+. \label{eq:eff_jump}
\end{align}
The relevant terms are discussed below.
They use the non-Hermitian Hamiltonian 
\begin{equation}\label{eq:nonhermitian}
    \hat{H}_\text{NH} = \hat{H}_e-\frac{i}{2 }\sum_k \hat{L}^\dag_k \hat{L}_k,
\end{equation}
to describe the dynamics of the excited states \citep{reiter_effective_2012}.
First we split our Hilbert space into ground and excited subspaces. The driving fields $\hat{V}$ are separated into excitation $\hat{V}_+$ and deexcitation $\hat{V}_-$ accordingly. Next, we calculate and invert the non-Hermitian Hamiltonian. Finally the effective operators are evaluated through a matrix product.
\subsection{Derivation of the effective decay of the OR gate}\label{ap:effopSpont}
In the following we calculate the effective operators that correspond to the couplings from Sec.\,\ref{sec:Setup}. The excited states are the internally excited state $\ket{e}$ and the harmonic oscillator excited state denoted as $\ket{1}_c$. The coupling $\hat{V}_+$ describes the weak excitation from the ground states
\begin{align}
    \hat{V}_+ = \frac{\Omega}{2} \left( \ketbraind{e}{1}{0} + \ketbraind{e}{2}{0}  \right).
\end{align}
The de-excitation is described by $\hat{V}_- = \hat{V}_+^\dag{}$. The decay processes relevant for the effective operators are those which couple excited and ground states. We consider spontaneous emission and oscillator decay
\begin{align}
    \hat{L}_{\gamma,j} = \sqrt{\gamma} \ketbraind{1}{j}{e}, \; \;  \hat{L}_{\kappa} = \sqrt{\kappa} \hat{a}.
\end{align}
The dynamics of the excited states are described by the non-Hermitian Hamiltonian from Eq.\,\ref{eq:nonhermitian}. It has block diagonal structure as only certain groups of excited states interact with each other. Two of these excited subspaces are shown in Fig.\,\ref{fig:desired} and Fig.\,\ref{fig:undesired}. To calculate the effective operators we invert the blocks individually. The first block describes excited-states $\left( \ket{e0},\ket{0e},\ket{00}\ket{1}_c \right)^T$. Using the complex detunings $\tilde{\delta}=\delta-\frac{i}{2}\kappa$ and $\tilde{\Delta}=\Delta-\frac{i}{2}\gamma$ its inverse is
\begin{align}
(\!\hat{H}_\text{nh}^1\!)^{-1}\!=\!\frac{1}{\tilde{\delta } \tilde{\Delta }^2\!-\!2g^2 \tilde{\Delta}}
\!\!\left(
\begin{array}{ccc}
 \!\tilde{\delta } \tilde{\Delta }\!-\!g^2 & g^2 & -\tilde{\Delta } g \!\\
 \!g^2 & \tilde{\delta } \tilde{\Delta}\!-\!g^2 & -\tilde{\Delta } g \!\\
\! -\tilde{\Delta } g & -\tilde{\Delta}  g & \tilde{\Delta }^2\!\\
\end{array}
\right)\! \!.
\end{align}
The second block describes the interaction shown in Fig.\,\ref{fig:desired}. It inverts to 
\begin{equation}
(\hat{H}_\text{nh}^2)^{-1}=\frac{1}{\tilde{\delta } \tilde{\Delta }_1-g^2}\left(
\begin{array}{cc}
 \tilde{\delta } & -g \\
 -g & \tilde{\Delta } \\
\end{array}
\right).
\end{equation}
In basis $\left( \ket{e1},\ket{01}\ket{1}_c\right)^T$. It is identical to that for states $\left( \ket{1e},\ket{10}\ket{1}_c\right)^T$. The forth and final block only contains the state $\ket{11}\ket{1}_c$. Its inverse is 
\begin{equation}
    (\hat{H}_\text{nh}^4)^{-1}= \frac{1}{\tilde{\delta}}.
\end{equation}
Using Eq. \eqref{eq:eff_jump} we can evaluate the effective jump operators. 
\begin{align}
    \hat{L}_{\text{eff}}^{\gamma_1}=\sqrt{\gamma}\frac{\Omega}{2} \biggl[\frac{1}{\tilde{\Delta}_{\text{eff},1}} &\ketbra{11}{01} +
    \frac{1}{\tilde{\Delta}_{\text{eff},2}}\ketbra{10}{00} \nonumber\\
    + \frac{1}{ g_{\text{eff},1}}   &\ketbra{10}{00} \biggl],
\end{align}

\begin{align}
      \hat{L}_{\text{eff}}^{\gamma_2}=\sqrt{\gamma}\frac{\Omega}{2}\biggl[  \frac{1}{\tilde{\Delta}_{\text{eff},1}} &\ketbra{11}{10}
      + \frac{1}{\tilde{\Delta}_{\text{eff},2}} \ketbra{01}{00} \nonumber\\
        +\frac{ 1}{g_{\text{eff},1}} &\ketbra{01}{00} \biggl].
\end{align}
Here $\tilde{\Delta}_{\text{eff},1} = \tilde{\Delta} -\frac{g^2}{\tilde{\delta}}$, $\tilde{\Delta}_{\text{eff},2} = \tilde{\Delta} - \frac{g^2}{\tilde{\delta}-g^2/\tilde{\delta}}$ and $g_{\text{eff},1} = -2\tilde{\Delta}+ \frac{\tilde{\delta}\tilde{\Delta}^2}{g^2}$. Summarizing terms we find the rate of the desired process $\ket{01}\rightarrow\ket{11}$:
\begin{align}
    \gamma_+ = \gamma \!\left(\!\frac{\Omega}{2}\!\right)^2 \! \abs{\frac{1}{\Delta_{\text{eff},1}}}^2= \gamma \!\left(\!\frac{\Omega}{2}\!\right)^2 \!\abs{\frac{\tilde{\delta}}{\tilde{\delta}\tilde{\Delta}-g^2}}^2.
\end{align}
The undesired decay process $\ket{00}\rightarrow\ket{10}$ occurs with the rate:
\begin{align}
    \gamma_- =& \gamma \left(\!\frac{\Omega}{2}\!\right)^2\abs{\frac{1}{\tilde{\Delta}_{\text{eff},2}}\!+\!\frac{1}{g_{\text{eff},1}}}^2
    \\= &\gamma \left(\!\frac{\Omega}{2}\!\right)^2\! \abs{\frac{\tilde{\delta}}{\tilde{\delta}\tilde{\Delta}-2g^2}}^2.
\end{align}
Photon loss in the cavity causes dephasing described by the following effective Lindblad operator
\begin{align}
\hat{L}_{\text{eff}}^{\kappa} = \sqrt{\kappa}\frac{\Omega}{2} & \biggl[ \frac{2}{ g_\text{eff,3}}\ketbra{00}{00} 
+ \frac{1}{ g_\text{eff,2}}\ketbra{10}{10}\nonumber\\
&+\frac{1}{ g_\text{eff,2}}\ketbra{01}{01}   \biggr].
\end{align}
Here $g_{\text{eff},2}= g-\frac{\tilde{\delta}\tilde{\Delta}}{g}$ and $g_{\text{eff},3} = 2g - \frac{\tilde{\delta}\tilde{\Delta}}{g}$.
The effective Hamiltonian obtained from Eq.\,\ref{eq:effhamiltonian} describes an AC-Stark shift of the driven levels.
\begin{align}\label{eq:effHo}
    \hat{H}_\text{eff}=   -\left(\!\frac{\Omega}{2}\!\right)^2&\Re{\frac{1}{\tilde{\Delta}_\text{eff,1} }} (\ketbra{01}{01}+\ketbra{10}{10}) \nonumber\\
    -\left(\!\frac{\Omega}{2}\!\right)^22&\Re{\frac{1}{\tilde{\Delta}_\text{eff,2} }} \ketbra{00}{00}. 
\end{align}
\subsection{Derivation of the effective operators of the oscillator-decay-based OR gate}\label{ap:OsziEffop}
The calculation for the oscillator-decay-based scheme is very similar to the one using spontaneous emission. The jump operators $\hat{L}_j^\gamma$,$\hat{L}^\kappa$ and the driving fields  $\hat{V}_j$ remain the same. Only the excited-state Hamiltonian is changed to
\begin{equation}
    \hat{H}_{e,j}= g_j(\hat{a}\ketbraind{e}{j}{1} + \hat{a}^\dag\ketbra{1}{j}{e}) + \Delta_j \ketbraind{e}{j}{e}.
\end{equation}{}
The non-Hermitian Hamiltonian keeps its block diagonal structure. What changes is which states couple to which block. Due to the different excited-state Hamiltonian $\hat{H}_\text{nh}^1$ now couples $\left( \ket{e1},\ket{1e},\ket{11}\ket{1}_c \right)^T$.
The second block, $\hat{H}_\text{nh}^2$, now couples  $\left( \ket{e0},\ket{10}\ket{1}_c \right) ^T$. States $\left( \ket{0e},\ket{01}\ket{1}_c \right) ^T$ couple to $\hat{H}_\text{nh}^3$. Due to the permutation in the couplings the states that couple through the effective operators are also permuted. 
\begin{align}
    \hat{L}^{\text{eff}}_{\gamma_1}=\sqrt{\gamma}\frac{\Omega}{2}\biggl[  \frac{1}{\tilde{\Delta}_{\text{eff},1}} &\ketbra{10}{00} +
    \frac{1}{\tilde{\Delta}_{\text{eff},2}}\ketbra{11}{01} \nonumber \\+  \frac{1}{ g_{\text{eff},1}} & \ketbra{11}{10} \biggl],
\end{align}
\begin{align}
      \hat{L}^{\text{eff}}_{\gamma_2}=\sqrt{\gamma}\frac{\Omega}{2}\biggl[ \frac{1}{ g_{\text{eff},1}} &\ketbra{11}{01} + \frac{1}{\tilde{\Delta}_{\text{eff},2}} \ketbra{11}{10} \nonumber\\
         +\frac{1}{\tilde{\Delta}_{\text{eff},1}} &\ketbra{01}{00} \biggl],
\end{align}
\begin{align}
       \hat{L}_{\text{eff}}^{\kappa}=-\sqrt{\kappa}\frac{ \Omega}{2}& \biggl[\frac{1}{ g_{\text{eff},1}} \ketbra{10}{00} +\frac{1}{ g_{\text{eff},1}}\ketbra{01}{00}\nonumber\\
    &+\frac{1}{g_{\text{eff},2}}\ketbra{11}{01}  +\frac{1}{g_{\text{eff},2}} \ketbra{11}{10}
     \biggl].
\end{align}
The effective Hamiltonian contains a shift of the driven levels.
\begin{align}\label{eq:effHoszi}
    \hat{H}_\text{eff}=   -\left(\!\frac{\Omega}{2}\!\right)^2&\Re{\frac{1}{\tilde{\Delta}_\text{eff,2} }} (\ketbra{01}{01}+\ketbra{10}{10})\nonumber\\
    -\left(\!\frac{\Omega}{2}\!\right)^22&\Re{\frac{1}{\tilde{\Delta}_\text{eff,1} }} \ketbra{00}{00}.
\end{align}
\section{Analytic optimization of the desired rate}\label{ap:detunings}
In order to optimize the detunings $\delta$ and $\Delta$ we can maximize the rate of the desired decay $\gamma_+$
\begin{equation}
    \gamma_+ = \gamma \!\left(\!\frac{\Omega}{2}\!\right)^2\abs{\frac{1}{\tilde{\Delta}_{\text{eff},1}}}^2=\gamma \!\left(\!\frac{\Omega}{2}\!\right)^2 \!\abs{\frac{\tilde{\delta}}{\tilde{\delta}\tilde{\Delta}-g^2}}^2.
\end{equation}
We rewrite the absolute value to show complex and real contributions
\begin{equation}
    \gamma_+ = \gamma \!\left(\!\frac{\Omega}{2}\!\right)^2 \frac{\delta^2+\frac{\kappa^2}{4}}{(\delta\Delta-g^2-\frac{\kappa \gamma}{4})^2+\frac{1}{4}(\delta\gamma + \Delta \kappa)^2}.
\end{equation}
The real contribution of the denominator is minimized by choosing
\begin{equation}
    \delta \Delta = g^2,
\end{equation}
neglecting the comparatively small $\kappa\gamma$ term. 
To optimize the ratio of the detunings $r=\frac{\Delta}{\delta}$ we further minimize the denominator $D(\delta,\Delta)$. We derive by $\Delta$ and $\delta$
\begin{align}
    \partial_\delta D(\delta,\Delta)=0 &\implies \delta = \frac{4g^2\Delta}{\gamma^2+4\Delta^2}, \label{eq:delta}\\
    \partial_\Delta D(\delta,\Delta)=0 &\implies \delta = \frac{g^2\pm \sqrt{g^4-\Delta^2\kappa^2}}{2\Delta}.
\end{align}
By setting the expressions equal and solving for $\Delta$ we find
\begin{equation}
    \Delta^\text{opt} = \pm \frac{\sqrt{\gamma(4g^2-\gamma \kappa)}}{2\sqrt{\kappa}} \approx \sqrt{\frac{\gamma}{\kappa}}g
\end{equation}
For the last equality we used that $\gamma \kappa \ll g^2$. By reinserting the expression for $\Delta $ into Eq.\,\ref{eq:delta} and using the same approximation we get:
\begin{equation}
    \delta^\text{opt} \approx \sqrt{\frac{\kappa}{\gamma}}g.
\end{equation}
The optimal ratio is:
\begin{equation}
    r = \frac{\Delta^\text{opt}}{\delta^\text{opt}}=\frac{\gamma}{\kappa}.
\end{equation}
This parameter choice minimizes the complex contribution to the denominator. Inserting the optimal detunings into the decay rate yields the maximal rate
\begin{align}\label{eq:gammaplusopt}
    \gamma_+^\text{opt} \approx \frac{\Omega^2}{4 \gamma}.
\end{align}
\section{Analytic time evolution of the OR gate}\label{ap:AnalyticEvo}
The incoherent flips of the ground states still make an analytic description challenging as they link all rate equations. We introduce additional simplifications:
The operation of the gate can be split into two main processes. The desired decay, shown in Fig.\,\ref{fig:desired}, is responsible for the gate action mapping $\ket{01}\rightarrow\ket{11}$. The undesired process $\ket{00}\rightarrow\ket{10}$, shown in Fig.\,\ref{fig:undesired}, produces errors. When the gate starts in the initial state $\ket{01}$ it would be best to wait for a long time such that most population is mapped to state $\ket{11}$. For initial state $\ket{00}$ one ideally would evaluate the result of the gate operation immediately as the error only increases in time. However, a gate should operate agnostically to the input state, meaning we cannot measure the input first and then determine the mode of operation. This gives rise to an optimal gate time $t_\text{opt}$ at which the average error $P_e^\text{avg}$ is minimal. 
The errors within the ground-states with rate $\gamma_g$ couple these evolutions. This makes an analytic analysis difficult. We can carefully untangle these processes allowing us to work with two separate and more manageable systems. The initial state $\ket{10}$ is resonantly mapped to $\ket{11}$, this does not produce any errors nor is it required for the gate action, therefore we can ignore it entirely for this analysis.

\begin{figure}
    \centering
    \includegraphics[width = \columnwidth]{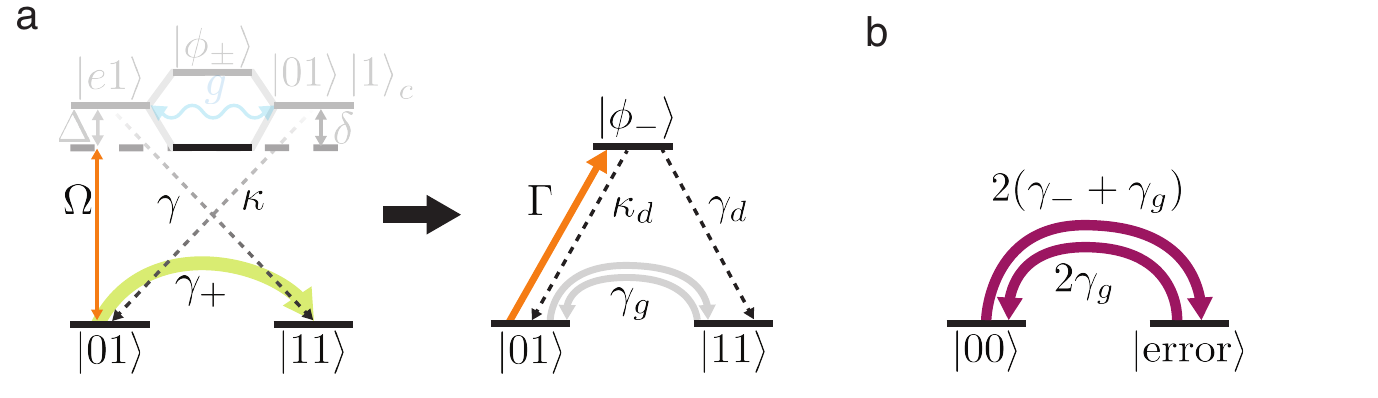}
    \caption{\textbf{Systems used for analytic description.} In Panel (a) we illustrate how the resonant process was modeled as a Lambda system with effective rates between the levels. Instead of the complex excited state structure, we only consider the lower energy dressed state $\ket{\phi_-}$. To further simplify the system the drive between ground-state $\ket{01}$ and excited-state $\phi_-$ is replaced by the rate $\Gamma$. The remaining decay rates remain unchanged. (b) For the undesired process, an even simpler model suffices. Initial state $\ket{00}$ decays to some erroneous state $\ket{\text{error}}$ with the rate $2(\gamma_-+\gamma_g)$. The decay in the opposite direction occurs with the smaller rate $2\gamma_g$.}
    \label{fig:AnalyticSystems}
\end{figure}

State $\ket{00}$ decays to $\ket{10}$ and $\ket{01}$ with effective rate $\gamma_-+\gamma_g$. If it ends up in $\ket{01}$ it is mapped towards $\ket{11}$ with the much stronger rate $\gamma_+$. We can model this as $\ket{00}$ decaying to some erroneous state $\ket{\text{error}}$ with rate $2(\gamma_-+\gamma_g)$. In the opposite direction only the ground-state flips occur with rate $2\gamma_g$. The rate equations are drawn in Fig.\, \ref{fig:AnalyticSystems} (b).
The undesired process can be described by the matrix equation
\begin{equation}
    \dot{\Vec{a}} = M_1 \Vec{a}.
\end{equation}
Where $\Vec{a}$ is a vector representation of the populations of $\ket{00}$ and $\ket{\text{error}}$. The matrix $M_1$ has the form
\begin{equation}
 M_1 = \left(
\begin{array}{cc}
 -2 \left(\gamma _-+\gamma _g\right) & 2 \gamma _g \\
 2 \left(\gamma _-+\gamma _g\right) & -2 \gamma _g \\
\end{array}
\right).
\end{equation}
To calculate the error probability of initial state $\ket{00}$ we exponentiate the matrix and calculate the scalar product with $\left(1,0\right)^T$. The population of the state $\ket{\text{error}}$ is the error probability
\begin{equation}
    P_e(00) = \frac{\gamma _-+\gamma _g}{\gamma _-+2 \gamma _g}(1-e^{-2 \left(2 \gamma _g+\gamma _-\right)t}).
\end{equation}

The resonant process is more difficult to model. The effective description is accurate as long as the assumptions of the formalism, $\Omega \ll g$ and $\Omega<\gamma$ are met. Numerically we find that the gate performs best at $\Omega\approx\gamma$. In this regime saturation and power broadening effects start to play a role. These effects are better described using the strong driving extension of the effective operator formalism \citep{reiter_scalable_2016}. The resonant rate is replaced by
\begin{equation}
    \gamma_+^{sd}= \frac{\gamma \Omega^2}{4(\gamma^2+2\Omega^2)}.
\end{equation}
In the weak-driving limit $\Omega\rightarrow0$ it reproduces the original rate.

We choose to model the resonant process as a single Lambda system. 
The states involved in the resonant process are $\ket{01}, \;\ket{11}$ and the dressed states $\ket{\phi_\pm}$ (Fig \ref{fig:AnalyticSystems}a). For our parameter choice the lower energy dressed state $\ket{\phi_-}$ is resonant with the drive while $\ket{\phi_+}$ is strongly detuned. This allows us to neglect the state $\ket{\phi_+}$. Now our system resembles a Lambda system with two decays and a drive. The coupling between $\ket{01}$ to $\ket{\phi_-}$ can be written as 
\begin{equation}
    \hat{V}_d = \frac{\Omega_d}{2}(\ketbra{01}{\phi_-}+\ketbra{\phi_-}{01}).
\end{equation}
The jump operators transform to
\begin{equation}
    \hat{L}_{\kappa _d}= \sqrt{\kappa _d}\ketbra{01}{\phi_-}, \; \hat{L}_{\gamma _d}= \sqrt{\gamma _d}\ketbra{11}{\phi_-}.
\end{equation}
For the optimal choice of detunings (Appendix \ref{ap:detunings}) the adjusted couplings are
\begin{equation}
    \Omega_d = \sqrt{\frac{\kappa}{\gamma+\kappa}}\Omega, \qquad \kappa _d=\gamma _d=\frac{\kappa\gamma}{\kappa+ \gamma}.
\end{equation}
The ground-states decay into each other with rate $\gamma_g$. Because of these ground-state decays and the presence of $\kappa^d$ describing this system analytically is still challenging. 
The next simplification we apply is replacing the Lambda system with three states and rates between them. 
We take as an ansatz for the decay rate from $\ket{01}$ to $\ket{\phi_-}$, that we denote with $\Gamma$\footnote{The expression for this rate was suggested by Tabea Bühler.}
\begin{equation}
\Gamma = \frac{\Omega_d^2 \gamma_{\text{tot}}}{\Omega_d^2 + \gamma_{\text{tot}}^2}=\frac{\Omega_d^2(\gamma _d+\kappa _d)}{\Omega_d^2+(\gamma _d+\kappa _d)^2}.
\end{equation}
Here $\gamma_{\text{tot}}$ stands for the total decay rate of state $\ket{\phi_-}$, which is $\gamma _d + \kappa _d$. The rate $\Gamma$ is motivated by the form of the rate that appears when we perform adiabatic elimination on a driven two-level system with one decay.
The decay rates $\kappa _d$, $\gamma _d$, and $\gamma_g$ remain unchanged. 
Now the system dynamics can be solved analytically.

We aim to calculate the optimal gate time $t_\text{opt}$ and the minimal average error $P_e^\text{avg}(t_\text{opt})$. First, we have to calculate the time evolution of initial states $\ket{01}$ and $\ket{00}$ in order to evaluate their error probabilities. To this end we follow the approximations illustrated in Fig.\,\ref{fig:AnalyticSystems}.

The resonant process is slightly more involved. Its evolution matrix is
\begin{equation}
   M_2 = \left(
\begin{array}{ccc}
 -\Gamma -\gamma _g & \text{$\kappa _d$} & \gamma _g \\
 \Gamma  & -\text{$\gamma _d$}-\text{$\kappa _d$} & 0 \\
 \gamma _g & \text{$\gamma _d$} & -\gamma _g \\
\end{array}
\right),
\end{equation}
for the population vector $\Vec{b}=\left(\ketbra{01}{01},\ketbra{\phi_-}{\phi_-},\ketbra{11}{11}\right)^T$. The eigenvalues evaluate to $\lambda_0=0$ and
\begin{align}
    \lambda_\pm =&\frac{1}{2} \Bigl(-\Gamma -2 (\text{$\gamma _d$}+\gamma _g)  \nonumber\\
    & \pm \sqrt{\Gamma ^2+4 \text{$\gamma _d$}^2+4 \gamma _g \left(\gamma _g-2 \text{$\gamma _d$}\right)}\Bigr) .
\end{align}
We inserted $\kappa _d = \gamma_d$. By exponentiation we find the error probability
\begin{align}\label{eq:error01}
    &P_e(01) = B_- e^{\lambda _- t}+B_+ e^{\lambda _+ t}\nonumber\\
    &+\frac{\Gamma  \text{$\gamma _d$}-\gamma _g \left(\Gamma +\gamma _g+\lambda _+ +\gamma _g+\lambda _-\right)}{\Gamma  \text{$\gamma _d$}+\Gamma  \gamma _g+4 \text{$\gamma _d$} \gamma _g},
\end{align}
with
\begin{align}
    B_\pm = \frac{\Gamma  \text{$\gamma _d$}-\gamma _g \left(\Gamma +2\gamma _g+\lambda _\mp \right)}{\Gamma  \text{$\gamma _d$}+\lambda _\pm \left(2 \Gamma +4 \text{$\gamma _d$}+4 \gamma _g\right)+\Gamma  \gamma _g+4 \text{$\gamma _d$} \gamma _g+3 \lambda _\pm^2}.
\end{align}
Equation \eqref{eq:error01} still does not allow us to calculate $t_\text{opt}$ as one ends up with a transcendental equation, making further approximations necessary.
Our initial assumption was $\gamma_g\ll\Omega \ll \gamma, \kappa$. By propagating this to the transformed rates we find 
\begin{equation}
    \gamma_g \ll\Gamma\ll \kappa _d,\gamma _d.
\end{equation}
Removing terms of size $\Gamma/\gamma _d$ and $\gamma_g/\gamma _d$ the eigenvectors can be approximated as
\begin{align}
    \lambda_+ &\approx-\frac{1}{2}(\Gamma +2\gamma_g),\\
    \lambda_- &\approx-\frac{1}{2}(\Gamma +\gamma_g+4\gamma _d).
\end{align}
Therefore, $\lambda_- \ll \lambda_+ < 0$. As a result the eigenvector to eigenvalue $\lambda_-$ decays far quicker than that of $\lambda_+$. For large times this allows us to remove it from the error probability.
We evaluate the average error using the approximation $P_e(10) ,P_e(11) \approx 0$.  
\begin{equation}
    P_e^\text{avg}=\frac{1}{4}(P_e(01) + P_e(00)).
\end{equation}
We derive by time and find
\begin{align}
    \frac{\partial}{\partial t}P_e^\text{avg} \approx \frac{1}{4}(2 \left(\gamma _-+\gamma _g\right) &e^{-2 \left(\gamma _-+2 \gamma _g\right)t}\nonumber\\
    +\lambda_+ B_+&e^{\lambda_+ t}).
\end{align}
Solving $\frac{\partial}{\partial t}P_e^\text{avg} =0$ for t we find the optimal gate time 
\begin{equation}
    t_\text{opt}= \frac{\log \left(-\frac{(\gamma_-+\gamma_g)}{B_+ \lambda_+}\right)}{2 \left(\gamma _-+2 \gamma _g\right)+\lambda _+}.
\end{equation}
If we insert the optimal gate time into the average error we find the minimal average error for a given set of parameters.

\begin{figure}
    \centering
    \includegraphics[width=\columnwidth]{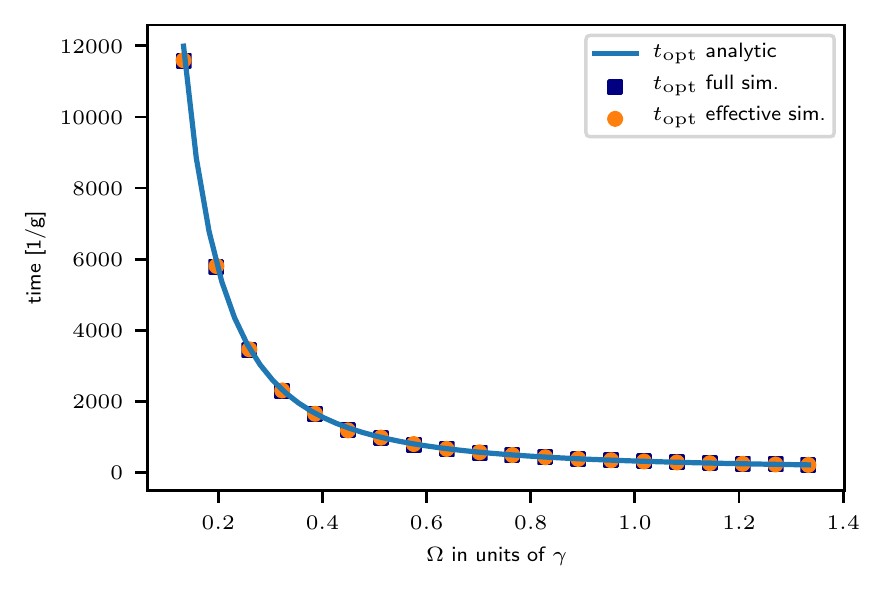}
    \caption{\textbf{Gate time.} Comparison of the analytic expression for $t_\text{opt}$ with the simulation of the effective operators and the full system. The analytic calculation fits the full simulation very well for a wide range of driving strengths. For weak driving the effective decay rate $\gamma_+ \propto \Omega^2$ (Eq.\,\ref{eq:gammaplusopt}). As a consequence decreasing $\Omega$ increases the gate operation time. This proportionality does not hold in strong driving regimes. There saturation and power broadening effects start to play a role. An increase in $\Omega$ no longer increases the rate. This causes the optimal gate time to decrease more slowly.}
    \label{fig:errorvstimeanalytic}
\end{figure}

In Fig.\,\ref{fig:errorvstimeanalytic} we compare the analytic expression for the optimal gate time with the simulation of the effective operators and the full system. Our analytic expression captures the dependence of gate time on the driving strength very well.
\begin{figure}
    \centering

    \includegraphics[width=\columnwidth]{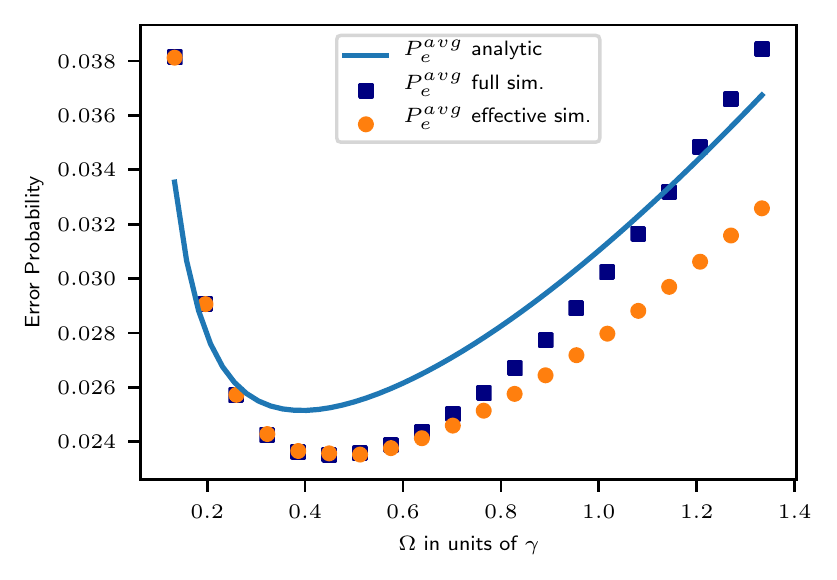}
    \caption{\textbf{Average error} Comparison of our analytic expression for the average error to simulations of the system for different $\Omega$. Weak driving makes the gate action slow and allows ground-state errors to occur. Strong driving disproportionately increases the rate of the undesired decay $\gamma_-$. Between these extremes there is an optimal value. The optimum driving strength from the analytic solution closely matches that found by simulations. Furthermore, the difference in minimal error probability is only $0.002$. For strong driving $(\Omega>\gamma)$ the analytic expression matches the full simulation better than the strong driving effective operators.}
    \label{fig:error}
 
\end{figure}
If we insert our expression for $t_\text{opt}$ into the average error $P_e^\text{avg}$ we get an expression for the error of a given set of parameters. In Fig.\,\ref{fig:error} we compare it to the simulations. Driving weakly increases the minimal average error as gate action is slow and ground-state errors occur. Driving strongly also increases the error. The rate of the resonant decay increases more slowly than the undesired decay due to saturation effects. This gives rise to an optimal driving strength $\Omega$. The optimum of the analytic calculation closely matches the full simulation. For the experimentally realistic parameters the analytic solution predicted a minimal average error at $3.0\%$ and optimal driving strength $\Omega=0.19$ GHz. The full simulation resulted in $P_e^\text{avg}=2.8\%$ at driving strength $\Omega=0.21$ GHz. In the limit of strong driving ($\Omega>\gamma$) the analytic solution better matches the full simulation than the simulation of the effective operators.

\begin{figure*}
    \centering
    \includegraphics[width=0.6\textwidth]{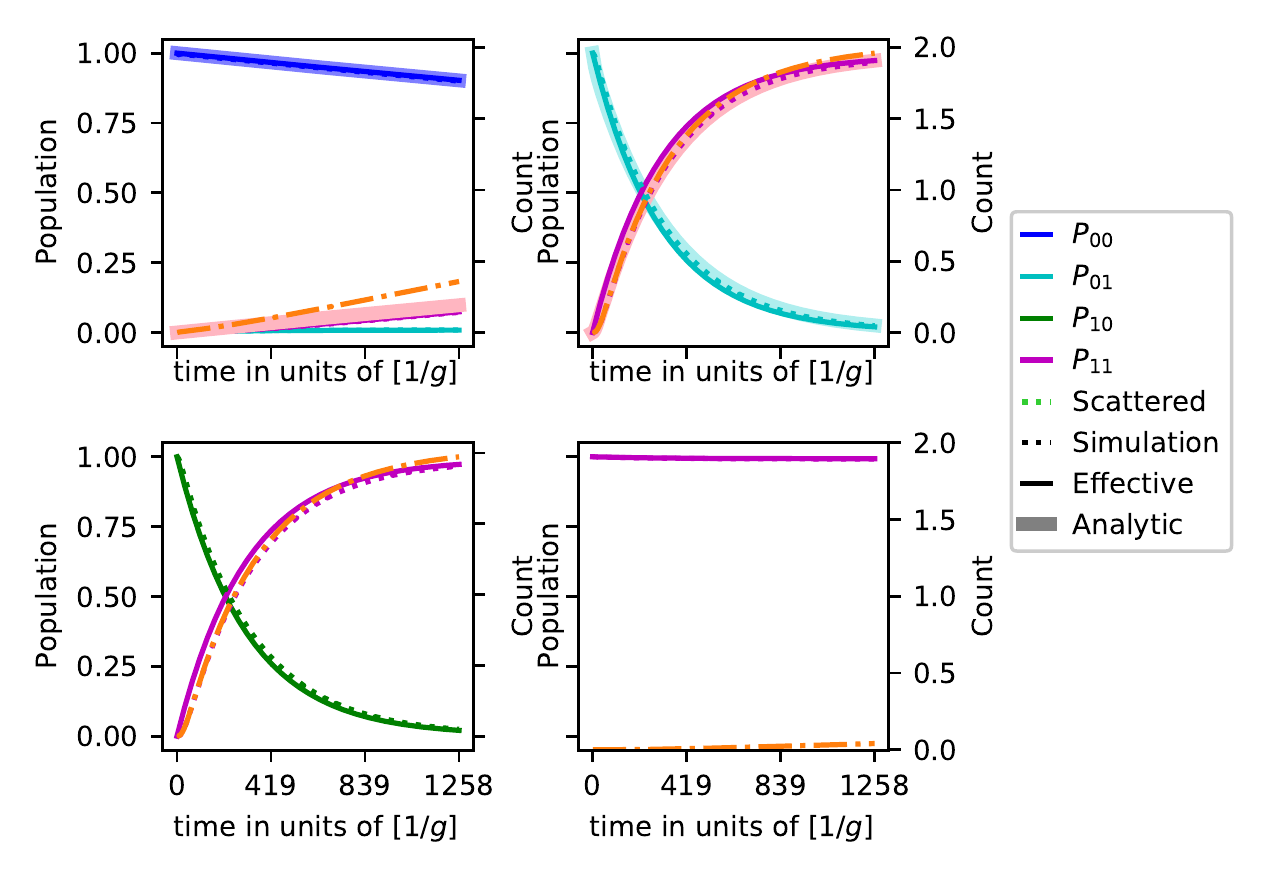}
    \caption{\textbf{Evolution of basis states and scattered photons.} The dotted lines show the full simulation, the solid line the simulation of the effective operators, and the thick opaque line the analytic solution. The colors denote the computational basis states. State $\ket{00}$ (top left) is slowly mapped to $\ket{01}$ and $\ket{10}$. These states are both resonantly mapped to $\ket{11}$. The analytic solution which ignores these intermediate states captures the dynamics of the system. During its evolution, it scatters 0.38 photons. State $\ket{01}$ (top right) is resonantly mapped to state $\ket{11}$. The dynamics are better described by our analytic solution than by the simulation of the effective operators. It scatters $1.96$ photons. As a consequence of the drive on the second qubit also state $\ket{10}$ is mapped to $\ket{11}$ (bottom left). Fortunately, this does not introduce any errors. As this process is not relevant for the error probability we do not plot the analytic solution but it should be identical to that of state $\ket{01}$. It also scatters 1.96 photons. In the bottom right plot, the evolution of state $\ket{10}$ is plotted. It is not addressed by a drive and therefore almost perfectly stable. The system parameters are  $(\frac{g}{2 \pi})=4.4$ GHz, $(\frac{\gamma}{2 \pi})=0.3$ GHz, $(\frac{\kappa}{2 \pi} )= 0.6$ GHz and $(\frac{\Omega}{2 \pi} )= 0.13$GHz.}
    \label{fig:Scattered4b4}
\end{figure*}

\section{Number of absorbed photons}\label{ap:Scattered}
Our gates operate by absorbing photons from a coherent source and then incoherently re-emitting these excitations. By estimating the number of absorbed photons we can judge the energetic cost of the operation. Classical gates operate irreversibly making them subject to Landauer's principle \cite{landauer_irreversibility_1961}. Future work could expand upon the consideration made here to put the energetic cost of our operations in context of Landauer's principle.

We can estimate the number of absorbed photons from the simple rate equations point of view introduced in Fig.\,\ref{fig:AnalyticSystems}. The analysis can be done analogously to that for optical pumping in Ref.\,\cite{happer_optical_1987}. The rate $\frac{dn}{dt} = (\gamma_d + \kappa_d) \bra{\phi_-}\rho(t)\ket{\phi-}$
describes how quickly photons are scattered or absorbed by the mirrors. To estimate the total number of scattered photons we integrate over this rate.
For our system, this consideration yields 2 photons scattered for initial states $\ket{01}$ and $\ket{10}$. States $\ket{00}$ and $\ket{11}$ are not excited and do not scatter any photons. Provided all inputs are equally likely we then scatter 1 photon on average. 

To verify this result we make use of numerical simulations. 
The operator
\begin{align}
    \sum_{\hat{L}_j \in S} \hat{L}_j^\dag{} \hat{L}j = \kappa \hat{a}^\dag{} \hat{a} +\gamma (\ketbra{e}{e} \otimes \mathbbm{1}+\mathbbm{1}  \otimes \ketbra{e}{e}),
\end{align}
evaluates how many excitations are emitted at a given moment. Here $S =\{ L_{\gamma_1},L_{\gamma_1},L_{\kappa}\}$. Integrating its expectation value over the time of operation yields the total number of emitted excitations.

In Fig.\,\ref{fig:Scattered4b4} the evolution of all four computational basis states is plotted with the number of scattered photons. Initial states $\ket{01}$ and $\ket{10}$ both scatter $1.96$ photons. Initial state $\ket{00}$ is only weakly excited, it therefore only scatters $0.38$ photons. State $\ket{11}$ is not addressed by the driving field. Only through a flip and subsequent excitation it can scatter photons. During operation it only scatters $0.02$ photons. Provided all input states are equally likely the gate scatters 1.08 photons on average.

\end{document}